\documentclass[twocolumn]{aastex62}
\pdfoutput=1 
\usepackage{amsmath,amstext}
\usepackage[T1]{fontenc}
\usepackage{apjfonts} 
\usepackage[figure,figure*]{hypcap}
\usepackage{subfigure}
\usepackage{epsfig}
\usepackage{longtable}
\usepackage{isotope}

\newcommand{\cutt}[1]{\textcolor{blue}{}}

\newcommand{\Ms}{{\ensuremath{{M}_{\odot} }}}
\newcommand{\Rs}{{\ensuremath{{R}_{\odot} }}}
\newcommand{\Zs}{\ensuremath{Z_\odot}}

\shorttitle{Pop III / EMP Binaries}
\shortauthors{Tsai et al.}

\begin{document}

\title{The Evolution of Population III and Extremely Metal-Poor Binary Stars}

\author{Sung-Han Tsai}
\affiliation{Department of Physics, National Taiwan University, Taipei 10617, Taiwan} 
\affiliation{Institute of Astronomy and Astrophysics, Academia Sinica, Taipei 10617, Taiwan} 

\author{Ke-Jung Chen}
\affiliation{Institute of Astronomy and Astrophysics, Academia Sinica, Taipei 10617, Taiwan} 

\author{Daniel Whalen}
\affiliation{Institute of Cosmology and Gravitation, University of Portsmouth, Dennis Sciama Building, Burnaby Road Portsmouth, PO1 3FX, United Kingdom}

\author{Po-Sheng Ou}
\affiliation{Department of Physics, National Taiwan University, Taipei 10617, Taiwan} 
\affiliation{Institute of Astronomy and Astrophysics, Academia Sinica, Taipei 10617, Taiwan} 

\author{Tyrone E. Woods}
\affiliation{National Research Council Herzberg Astronomy \& Astrophysics, 5071 West Saanich Rd, Canada}

\begin{abstract}

Numerical simulations have now shown that Population~III (Pop~III) stars can form in binaries and small clusters and that these stars can be in close proximity to each other.  If so, they could be subject to binary interactions such as mass exchange that could profoundly alter their evolution, ionizing UV and Lyman-Werner (LW) photon emission and explosion yields, with important consequences for early cosmological reionization and chemical enrichment.  Here we investigate the evolution of Pop~III and extremely metal-poor binary stars with the \texttt{MESA} code.  We find that interactions ranging from stable mass transfer to common envelope evolution can occur in these binaries for a wide range of mass ratios and initial separations.  Mass transfer can nearly double UV photon yields in some of these binaries with respect to their individual stars by extending the life of the companion star, which in turn can enhance early cosmological reionization but also suppress the formation of later generations of primordial stars.  Binary interactions can also have large effects on the nucleosynthetic yields of the stars by promoting or removing them into or out of mass ranges for specific SN types.  We provide fits to total photon yields for the binaries in our study for use in cosmological simulations.

\end{abstract}

\keywords{binary stars --- stellar evolution --- reionization --- early universe --- Population~III stars --- primordial galaxies}

\section{Introduction}

The birth of Population~III (Pop~III) stars at $z \sim$ 20 - 25 marked the end of the cosmic Dark Ages and the onset of cosmological reionization \citep{wan04,ket04,abs06,awb07}.  Pop~III stars were also the first great nucleosynthetic engines in the universe, forging the heavy elements required for the later formation of planets and life \citep{get07,wet08a,jet09b,jw11,mag20,latif20c}. Pop~III stars were originally thought to form in isolation, one per halo \citep{bcl99,abn02} but were later found to occur in binaries and small multiples in simulations \citep[e.g.,][]{turk09,stacy10,clark11,get12,susa13,susa14,susa19,sug20,latif20a,pat21a,latif21a}.  Although the Pop~III IMF remains uncertain it is thought to be top heavy, with masses ranging from a few tens to hundreds of solar masses \citep{hir15,latif22a}.

The existence of massive Pop~III binaries raises the possibility of mass transfer that could radically alter the properties of both stars, along with their ionizing photon emission rates \citep{chen15} and elemental yields if they die as supernovae (SNe).  Their evolution begins with two stars on the main sequence, with one later growing to a radius that leads to mass transfer to its companion.  Spectroscopic studies have shown that most massive stars in the Universe today have a companion which is close enough to exchange mass at some point in their evolution \citep{ragh10,astlux}.  Besides having a potentially important impact on early cosmological reionization and chemical enrichment, mass exchange in primordial binaries may have governed gravitational wave (GW) production in the early Universe by setting the masses and inspiral times of the black holes \citep[BHs; e.g.,][]{kin14,in16,til16,lb20}.

The evolution of single Pop~III stars has been examined in a number of studies \citep{hw02,hw10,yoon12,murphy21}. Modelling the evolution of Population III {\it binaries}, however, has to-date largely focussed on a few isolated systems \citep[e.g.,][]{lyj08, song20} or on rapid calculations using interpolated fits to single star evolution to model binary populations \citep[e.g.,][]{Inayoshi2017,Hijikawa21,Sartorio23}.  Here, we investigate the evolution of interacting binary stars at zero and very low metallicities, like those found in the primordial universe and the first galaxies, using a stellar evolutionary code to follow the evolution of these stars and their response to mass tranfer. This allows us to examine their mass transfer, final masses and ionizing UV luminosities over their lifetimes in detail.  Our models evolve the stars from the zero-age main sequence to collapse for a grid of total masses, mass ratios, metallicities, and separations.  In Section 2 we describe our numerical models and in Section 3 we discuss the evolution and properties of the stars. In Section 4, we analyze the effects of mass transfer on the stars and discuss radiative feedback from binaries at high redshifts in Section 5.  We conclude in Section 6.

\section{Numerical Method}

We use the \texttt{MESA} code to evolve the binary stars in our study.  \texttt{MESA} is a one-dimensional (1D) Lagrangian stellar evolution code with convective mixing and nuclear burning that is implicitly coupled to updates of the equations of stellar structure \citep[version r10108;][]{paxt11,paxt13,paxt15}. We consider three metallicities, $Z$, where $Z = $ 0, 10$^{-3}$ \Zs, or 10$^{-2}$ \Zs\ \citep[where \Zs\ $=$ 0.017 as in][]{grev96}. For each metallicity we sample 10 donor star masses, $M_1$, from 10 - 100 \Ms\ in 10 \Ms\ increments. The mass of the companion star, $M_2$, is determined by two ratios, $q_2 = M_2/M_1 =$ 0.8 and 0.5. These mass combinations yield 60 binaries.  For each binary we consider 10 initial binary separations, $a$, from 100 to 1000 $R_{\sun}$ in 100 $R_{\sun}$ increments for a total of 600 models.

\subsection{Model Setup}

We initialize the donor and companion stars separately as fully-convective $n =$ 3 polytropes with the specified masses.  We set their initial helium mass fractions, $Y$, so that they increased linearly from $Y = 0.2477$ \citep{peimbert07} to $Y = 0.28$ 0ver the interval $Z =$ 0 - \Zs\, although the maximum metallicity we considered was 0.01 \Zs. The hydrogen mass fractions, $X$, then adjusted automatically to satisfy mass conservation law, $X+Y+Z=1$.  After initialization, the protostars begin to contract under their own gravity. Their core temperatures then rise and they eventually begin nuclear burning.  However, the onset of p - p burning does not produce enough energy to halt contraction in Pop III protostars so it continues until the triple alpha chain builds up enough C to begin the CNO cycle.  CNO burning halts contraction and the star enters the main seqence.  EMP protostars reach the main sequence earlier because the CNO cycle begins nearly immediately in them.

\texttt{MESA} adaptively rezones the stars during their evolution. Mesh refinement is triggered when gradients in pressure, temperature and $^4$He abundances between adjacent zones exceed certain values: $\delta \log P/P >$ 1/30, $\delta \log T/T >$ 1/80, and $\delta \log(\chi + 0.01) >$ 1/20 $\log \chi$, where $\chi$ is the $^4$He mass fraction \citep[section 6.4 of][]{paxt11}.  This prescription results in larger numbers of zones typically being allocated near the center of the star to resolve nuclear burning and convective processes and the stars being partitioned into 1000 - 3000 mass zones.  In this first stage we use the 8-isotope \texttt{basic.net} nuclear reaction network (H, \isotope[3]{He}, \isotope[4]{He}, \isotope[12]{C}, \isotope[14]{N}, \isotope[16]{O}, \isotope[20]{Ne} and \isotope[24]{Mg}), which includes the pp chain, CNO cycle, and helium burning.  Later, when the stars reach carbon burning \texttt{MESA} switches to the 9-isotope \texttt{co\_burn.net} network, which adds in \isotope[28]{Si}, and then uses the APPROX21 network when the star reaches oxygen burning.  We exclude mass loss due to stellar winds and pulsations because they are thought to be negligible at our metallicities \citep{vink01,bhw01}.  

We use the Ledoux criterion for convection, which is modeled with Henyey's mixing-length theory, and we set the mixing-length scaling parameter to 1.5 \citep{henyey1965}. The equation of state (EOS) in our models is a composite of several datasets, including the OPAL/SCVH tables \citep{rogers02,saumon1995}, which are used at lower densities and temperatures like those in the outer regions of the star and its atmosphere, and the \texttt{HELM} and \texttt{PC} EOSs \citep{ts00,potekhin10}, which are used hotter and denser regions, like those in the core of the star \citep[see Figure 1 and Section 4.2 of][]{paxt11}. All \texttt{MESA} files required to reproduce our results are available at \dataset[doi:10.5281/zenodo.7949967]{https://doi.org/10.5281/zenodo.7949967}.

\subsection{Binary Evolution}

We then use \texttt{MESAbinary} \citep{mad06,lin11,paxt15} to co-evolve the two ZAMS stars as a binary.  We evolve the binary until the donor star forms a collapsing iron core or the binary becomes a common envelope candidate.  Mass exchange between the stars begins when the donor star begins to overflow its Roche lobe, whose radius, $R_{\mathrm{RL,1}}$, can be approximated as \citep{egg83}
\begin{equation}\label{key}
\frac{R_{\mathrm{RL,1}}}{a}=\dfrac{0.49q_1^{2/3}}{0.6q_1^{2/3}+\ln(1+q_1^{1/3})},
\end{equation}
where $q_{1} = M_{1} / M_{2}$.
We use the Ritter scheme \citep{ritt88} to model mass transfer between binary stars in \texttt{MESAbinary} because it is more stable, especially if an extended atmosphere of a star triggers Roche-lobe overflow.  The mass transfer rate is
\begin{equation}
\dot{M} = -{\dot{M}}_{0} \, \mathrm{exp} \left(\frac{R_{1} - R_{\mathrm{RL,1}}}{H_{\mathrm{\mathrm{P,1}}}/\gamma(q_2)}\right),
\label{eq:xfer}
\end{equation}
where $H_{\mathrm{P,1}}$ is the pressure scale height at the photosphere of the donor star and
\begin{equation}
{\dot{M}}_{0} = \frac{2\pi}{e^{1/2}} F_{1}(q_2)\frac{R^{3}_{\mathrm{L}}}{GM_{1}}\left(\frac{k_{\mathrm{B}}T_{\mathrm{eff}}}{m_{\mathrm{p}}\mu_{\mathrm{ph}}}\right)^{3/2}\rho_{\mathrm{ph}},
\end{equation}
where $k_{\mathrm{B}}$ is the Boltzmann constant, $m_{\mathrm{p}}$ is the proton mass, $T_{\mathrm{eff}}$ is the effective temperature of the donor, $\mu_{\mathrm{ph}}$ and $\rho_{\mathrm{ph}}$ are the mean molecular weight and density at its photosphere, and $F_1(q_2)$ and $\gamma(q_2)$ are fitting functions \citep[equations 15 and 16 in][]{paxt15}. The eccentricities of the orbits of the stars in our models are zero so their trajectories in the binary are two concentric circles. When mass transfer occurs, their separation evolves because of angular conservation as
\begin{equation}\label{eqseparation}
a\arcmin = \frac{M_{1}^2M_{2}^2}{(M_{1}-dm)^2(M_{2}+dm)^2}a,
\end{equation}
where $a\arcmin$ is the new separation after a mass transfer of $dm$.

\subsection{Accretion Physics} \label{subsection:4}


\begin{figure}
\centering
\includegraphics[width=0.45\textwidth]{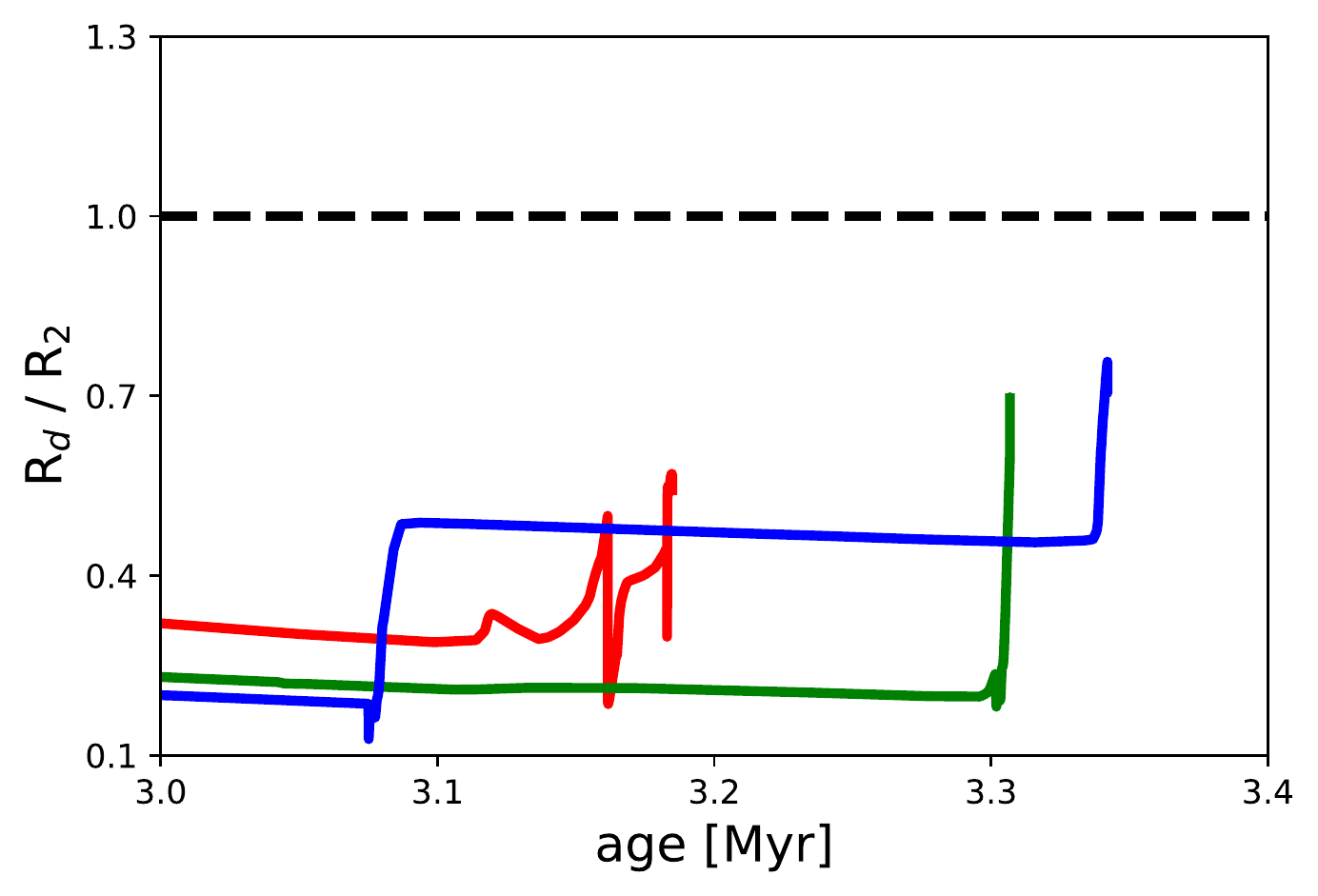}
\label{fig:diskevol}
\caption{$R_{\mathrm{d}}/R_2$ for binaries with a total mass of 90 \Ms, $q_2 =$ 0.5 and $a =$ 20 $R_1$.  Blue:  $Z = $ 10$^{-2}$ \Zs; green:  $Z = $ 10$^{-3}$ \Zs.}
\end{figure}


\begin{figure*}
\centering
\includegraphics[width=1\textwidth]{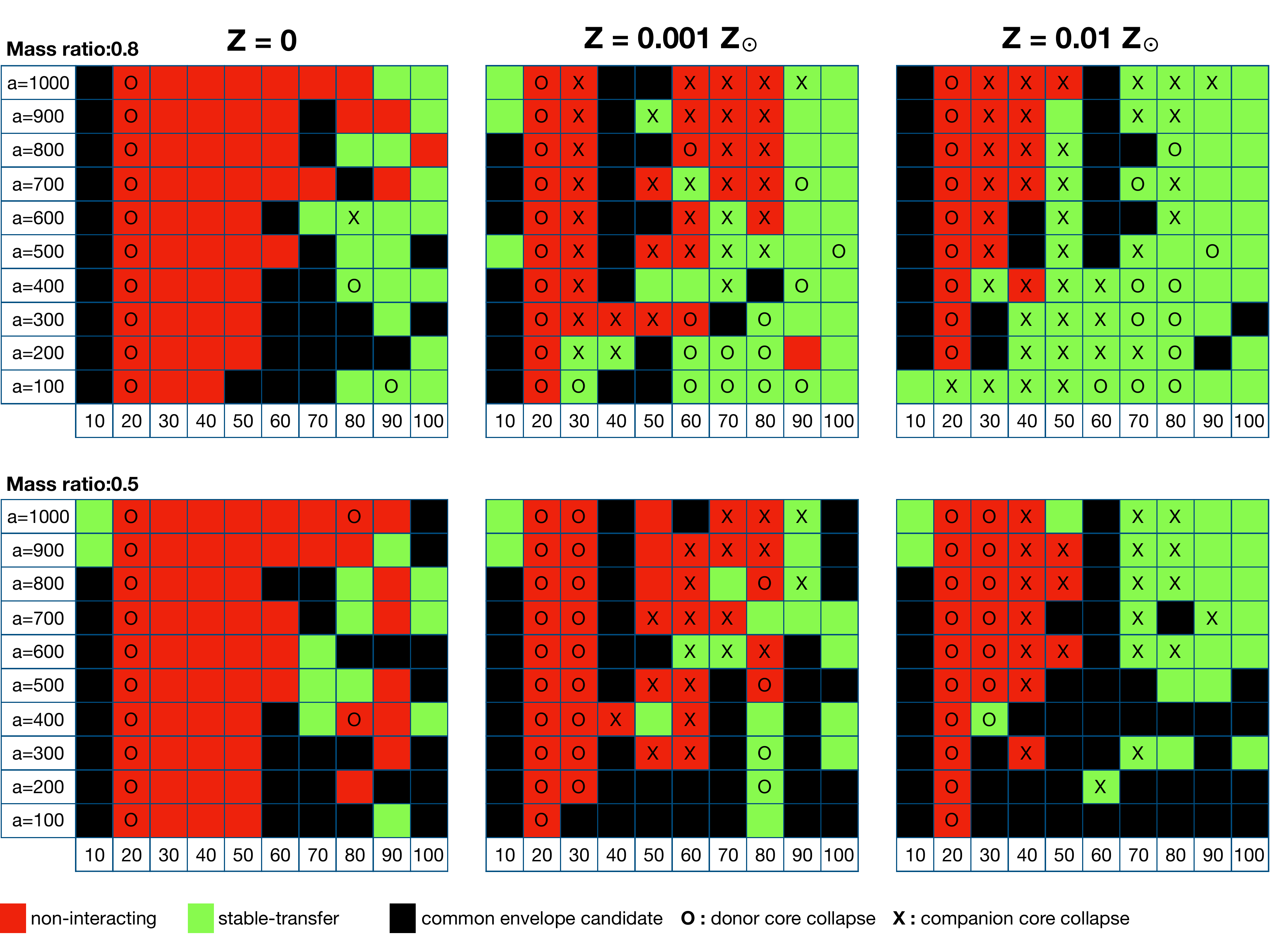}
\label{fig:data_base}
\caption{Types of mass transfer between the binary stars: non-interacting (red), stable-transfer (green), and common envelope evolution (black).  Models that could only be evolved to the collapse of the donor star because of numerical issues are marked by circles.  Those that could be evolved up to the collapse of the companion star are marked by X's.}
\end{figure*}

Mass transfer can create an accretion disk around the companion star that converts the kinetic energy of the inflow into thermal and kinetic energy at the surface of the star, which either heats the star or is released as radiation.  Thus, in principle the properties of the disk are needed to model the evolution of the star and its ionizing UV luminosities.  Gas from the donor falls from $L_{1}$ into the gravitational field of the second star and, depending on its angular momentum, orbits the star at a radius, $R_{\mathrm{d}}$, the Keplerian orbit that has the same angular momentum as the gas at $L_{1}$. The velocity of the gas at $R_{\mathrm{d}}$ is
\begin{equation}\label{eq4}
v_{d} = \left(\frac{GM_{2}}{R_{\mathrm{d}}}\right)^{1/2}.
\end{equation}
Because the gas at $L_{1}$ and $R_{d}$ has the same angular momentum, 
\begin{equation}\label{eq5}
R_{\mathrm{d}}v_{\mathrm{d}}=b^{2}\omega,
\end{equation}
where $b$ is the distance between $L_{1}$ and the center of the second star and $\omega$ is the angular velocity of the gas.  Using $\omega=2\pi/P$, where $P$ is the binary period, and Kepler's third law, $R_{d}$ becomes
\begin{equation}\label{eq6}
R_{\mathrm{d}}/a=(1+q)(b/a)^{4}.
\end{equation}

If the ratio of the radius of the disk and companion star is less than one ($R_{\mathrm{d}}/R_{2}<1$), the gas will fall directly onto the star without forming a disk.  If this ratio is greater than one, a disk forms.  In our binary models accretion disks rarely form, and if they do their radii are too small to be described using simple disk models or they form late in the life of the star, as shown for two metallicities in Figure~\ref{fig:diskevol}.  We therefore ignore effects due to disks in our models and assume the infall energy is radiated away, with no effect on the surface temperature of the star.


\begin{table}
\vspace{0.3in}
\begin{center}
\begin{tabular}{ |c|c|c|c| } 
\hline
     & $Z = $ 0 & $Z = $ 0.001 \Zs\ & $Z = $ 0.01 \Zs\ \\ 
\hline
non-interacting        &  54\%     &  39\%     & 23.5\%  \\
stable-transfer         &  16.5\%  &  29.5\%  & 42.5\%   \\ 
common envelope  &  29.5\%  &  31.5\%   & 34\%      \\ 
\hline
\end{tabular}
\end{center}
\caption{Percentages of non-interacting, stable-transfer and overflowing binaries by metallicity.}
\label{tbl:1}
\end{table}

\subsection{Ionizing UV Luminosities}

We calculate ionizing photon emission rates for the stars by treating them as black bodies with spectral radiance densities $B_{\nu}$:
\begin{equation} \label{key}
B_{\nu}(T)=\dfrac{2h\nu^{3}}{c^{2}}\frac{1}{e^{h\nu/k_{\mathrm{B}}T}-1}.
\end{equation}
Assuming $T$ to be the effective temperature $T_{\mathrm{eff}}$ of the star, the ionizing photon flux per unit solid angle per unit time is
\begin{equation}\label{key}
f_{i}=\int_{\nu_{\mathrm{min}}}^{\infty}\frac{\pi B_{\nu}(T)}{h\nu}d\nu,
\end{equation}
where $h\nu_{\mathrm{min}}$ is the ionization energy threshold of H (13.6 eV), He~I (24.6 eV), or He~II (54.4 eV).  Lyman-Werner (LW) fluxes with $h\nu_{\mathrm{min}}= $ 11.2 eV are calculated in the same manner.  The total ionizing photon emission rate $\dot{N}$ is then
\begin{equation}\label{key}
\dot{N}=4\pi R^{2}f.
\end{equation}
Because radiation from the accretion disk is excluded in our models, the ionizing photon emission rate of the binary is just the sum of the rates of its stars:
\begin{equation} \label{key}
\dot{N}=4\pi (R_{1}^{2}f_{1}+R_{2}^{2}f_{2}),
\end{equation}
where $R_{1}$ and $R_{2}$ are the radii of the donor and companion stars, respectively. This rate evolves during mass transfer as the effective temperatures of the stars change.

\section{Results}


\begin{figure*}
\centering
\includegraphics[width=\textwidth]{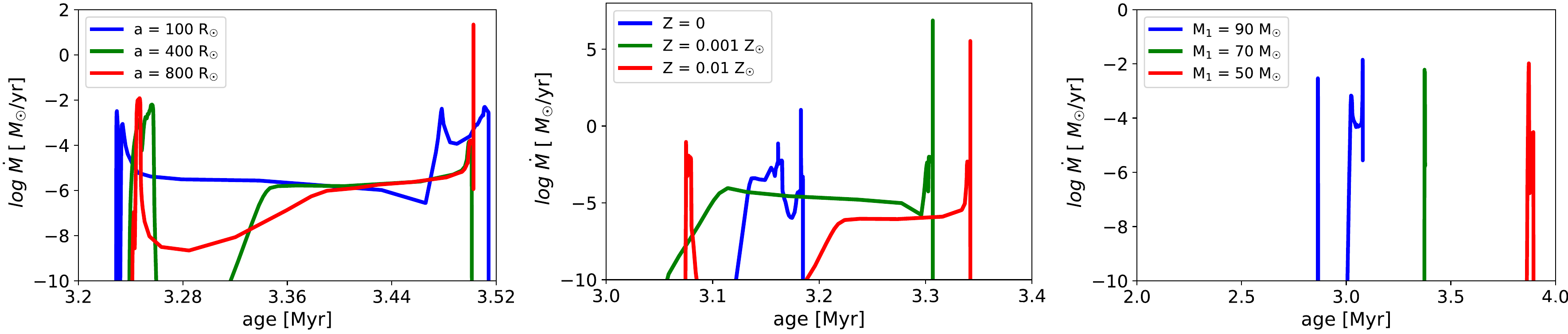}
\label{fig:transfer}
\caption{Mass transfer rates for nine of the binaries.  Left: three binaries with total masses of 80 \Ms, $q_{2} =$ 0.8, $Z = $ 10$^{-2}$ \Zs\ and $a =$ 100, 400 and 800 \Rs.  Middle:  three binaries with donor masses of 90 \Ms, $q_{2} =$ 0.8, $a =$ 100 \Rs, and $Z =$ 0, 10$^{-3}$ \Zs\ and 10$^{-2}$ \Zs.  Right: three binaries with $q_{2} =$ 0.8, $a =$ 500 \Rs, $Z =$ 10$^{-2}$ \Zs,  and donor masses of 90 \Ms, 70 \Ms, and 50 \Ms. Mass transfer generally proceeds for $\sim$ 0.1 - 0.3 Myr at 2.5 - 4 Myr, after the more massive donor stars exit the main sequence.}
\end{figure*}

The binaries in our study evolve along three pathways: non-interacting, stable mass transfer, and common envelope evolution, as summarized in Figure~\ref{fig:data_base} and Table~\ref{tbl:1}.  When the donor star exits the main sequence, hydrogen shell burning expands its outer envelope.  Large separations or low metallicities prevent mass transfer if the stars are too far away from one another or are too compact for Roche lobe overlap. Stable mass exchange occurs when the radius of the donor exceeds its Roche lobe and it transfers mass to its companion.  If the donor star transfers gas to to its companion too quickly, the companion can expand beyond its own Roche lobe and the two stars can come into contact. Gas then envelopes both stars, and they enter a common envelope phase that leads to much higher and more irregular transfer rates.  Stable mass transfer can later lead to common envelope evolution because the exchange can change the structures of the stars and their separation.  

61.2\% of our 600 binaries exhibit some type of mass exchange, either stable transfer (29.5\%) or common envelope evolution (31.7\%).  The number of interactions increases with mass at all metallicities because the stars have larger radii.  Only 54\% of Pop~III binaries exchange mass because the stars are more compact.  In contrast, 61\% of the $Z =$ 10$^{-3}$ \Zs\ binaries and 76.5\% of the $Z =$ 10$^{-2}$ \Zs\ binaries transfer mass.  Metallicity plays a key role in binary interactions because those with higher metallicities form puffier stars that more easily initiate mass transfer later in their lives.  As noted in Figure~\ref{fig:data_base}, not all of the binaries could be evolved up to the death of the donor star because of numerical instabilities that arose in the \texttt{MESA} runs, and even fewer could be evolved up to the collapse of the comapnion star, about one in six.  This latter fact was due to the difficulties in evolving the companion after significant fractions of the mass of the donor were deposited onto its outer layers.  Nevertheless, it was possible in most cases to capture the effects of mass transfer on both stars as it occurred.

\subsection{Stable Mass Transfer}

Stable mass transfer occurs in 16.5\% of the Pop~III binaries, 29.5\% of the 10$^{-3}$ \Zs\ binaries, and 42.5\% of the 10$^{-2}$ \Zs\ systems. It favors binaries with smaller separations and higher donor masses because small separations create compact Roche lobes that are easier to fill and donors with large masses have larger radii at late times. In Pop~III binaries, 20 - 50 \Ms\ donor stars do not expand enough to fill their Roche lobes before core collapse.  However, 10 \Ms\ donor stars with can exceed their Roche lobes during  sudden pulsations of their envelopes during helium burning in our simulations. This clustering holds for 10$^{-3}$ \Zs\ binaries but over a greater range in total mass at the low and high ends and in separation, and these ranges are even broader in the 10$^{-2}$ \Zs\ models. In our models, stable transfer mostly happens when $q_2 =$ 0.8.

\subsection{Common Envelope Evolution}

Mass exchange in binary stars with lower mass ratios usually leads to common envelope evolution because they have larger Roche lobes with higher transfer rates that are more likely to lead to companion star overflow.  Like the stable transfer cases, the common envelope candidates are clustered at small separations and the highest and lowest masses at zero metallicity.  Both types of mass transfer occur primarily in donor stars that are either $\gtrsim$ 40 \Ms\ or $\lesssim$ 10 \Ms.  37.2\% of the donor stars in our study reached core collapse but we had to terminate 31.2\% of the runs because of small time steps.  However, most of these latter cases nearly reached core collapse and emitted most of their ionizing photons. 


\begin{figure} 
\begin{tabular}{c}
\includegraphics[width=0.45\textwidth]{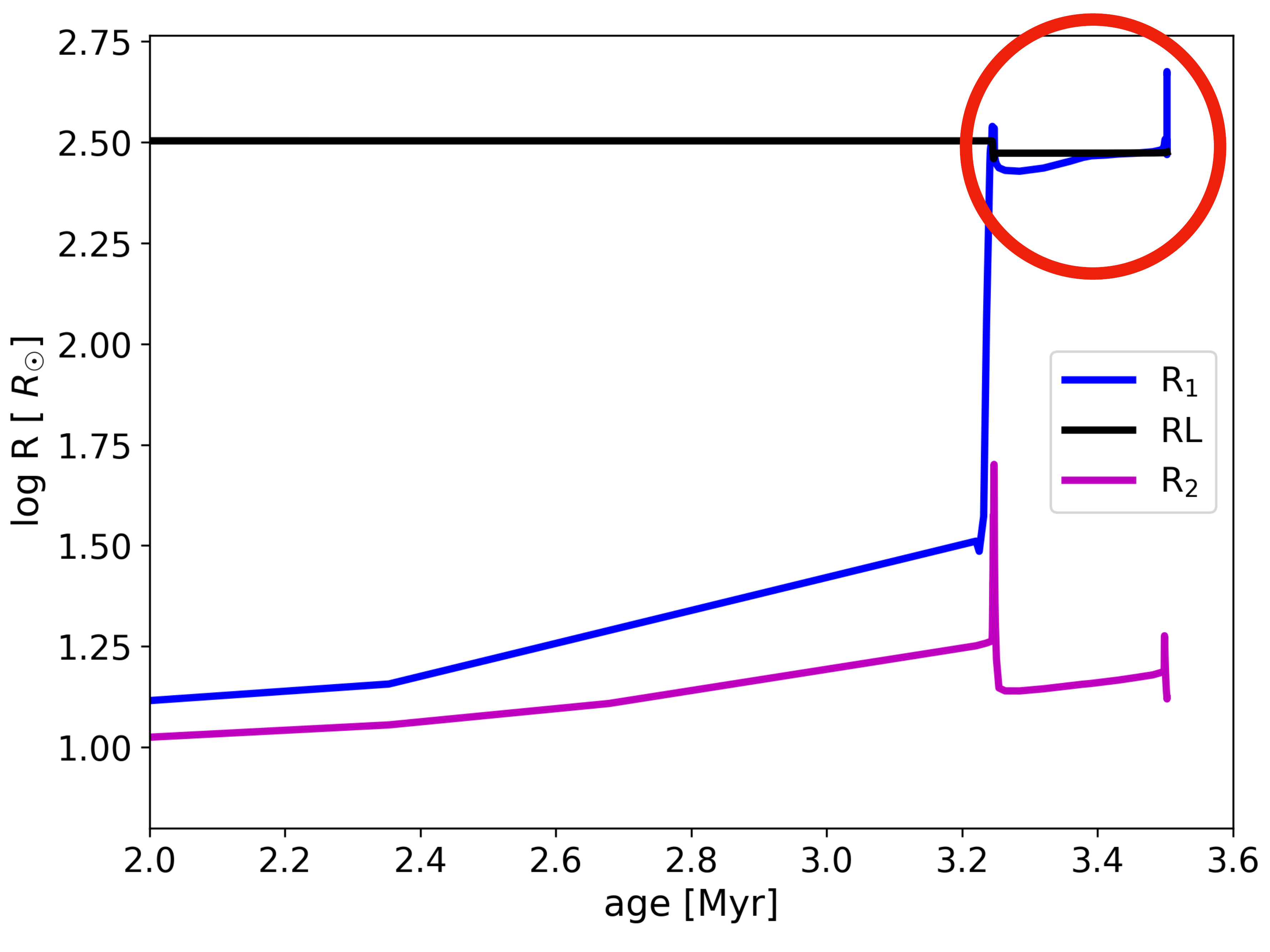} \\
\includegraphics[width=0.45\textwidth]{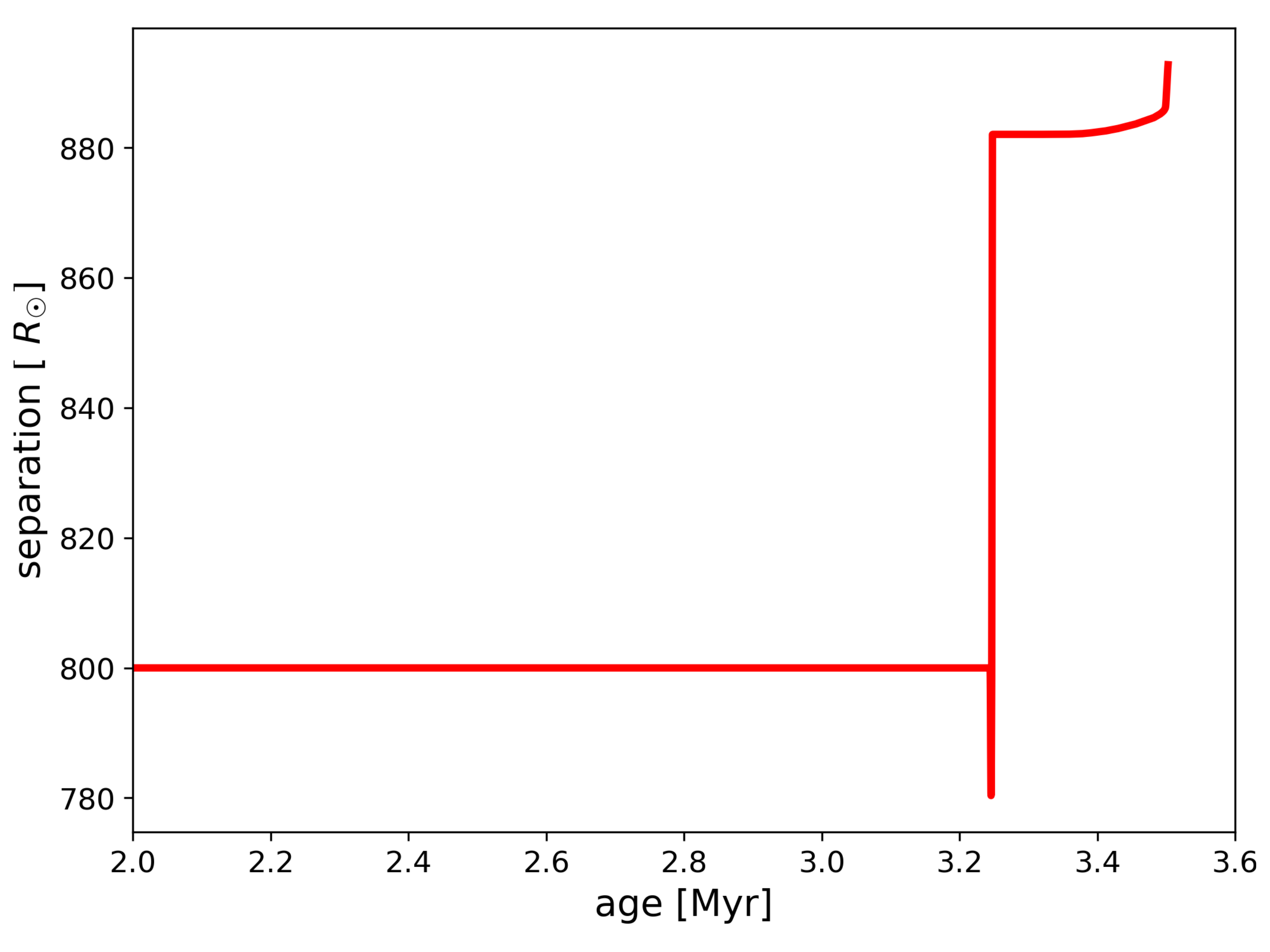}
\end{tabular} 
\label{fig:radius}
\caption{Top: Evolution of stellar radii in a binary with a donor mass of 80 \Ms, $q_2 = $ 0.8, $Z = $ 10$^{-2}$ \Zs\ and initial separation $a = $ 800 $R_{\sun}$.  "RL" is the Roche lobe radius of the donor star.  Bottom:  separation between the stars over time, which abruptly decreases from 800 to 780 \Rs\ when they begin to transfer mass and then jumps to 880 \Rs, after which  transfer ends.  The red circle marks the time interval over which mass transfer occurs.}
\end{figure}

\subsection{Mass Transfer Rates}


\begin{figure*}
\centering
\includegraphics[width=\textwidth]{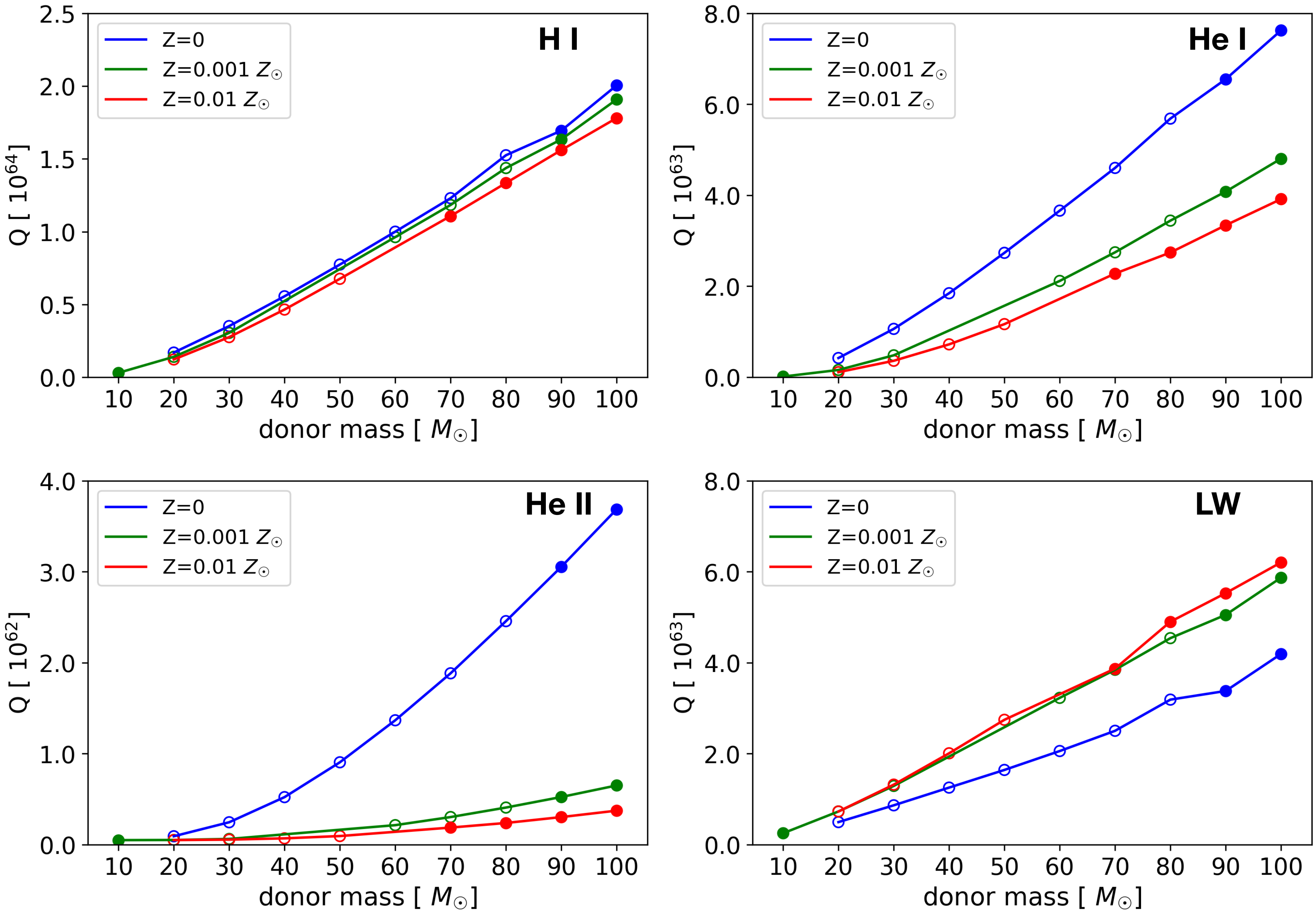}
\label{fig:ioniz}
\caption{Total UV photon yields, $Q$, for H I, He I, and He II ionizations and the LW band as a function of donor mass and metallicity for $a =$ 1000 \Rs\ and $q_{2} =$ 0.8.  The filled and empty circles denote stable transfer and non-interacting binaries, respectively. UV photon yields vary inversely with metallicity but LW photon yields increase with metallicity.}
\end{figure*}
\par


\begin{figure*}
\centering
\includegraphics[width=\textwidth]{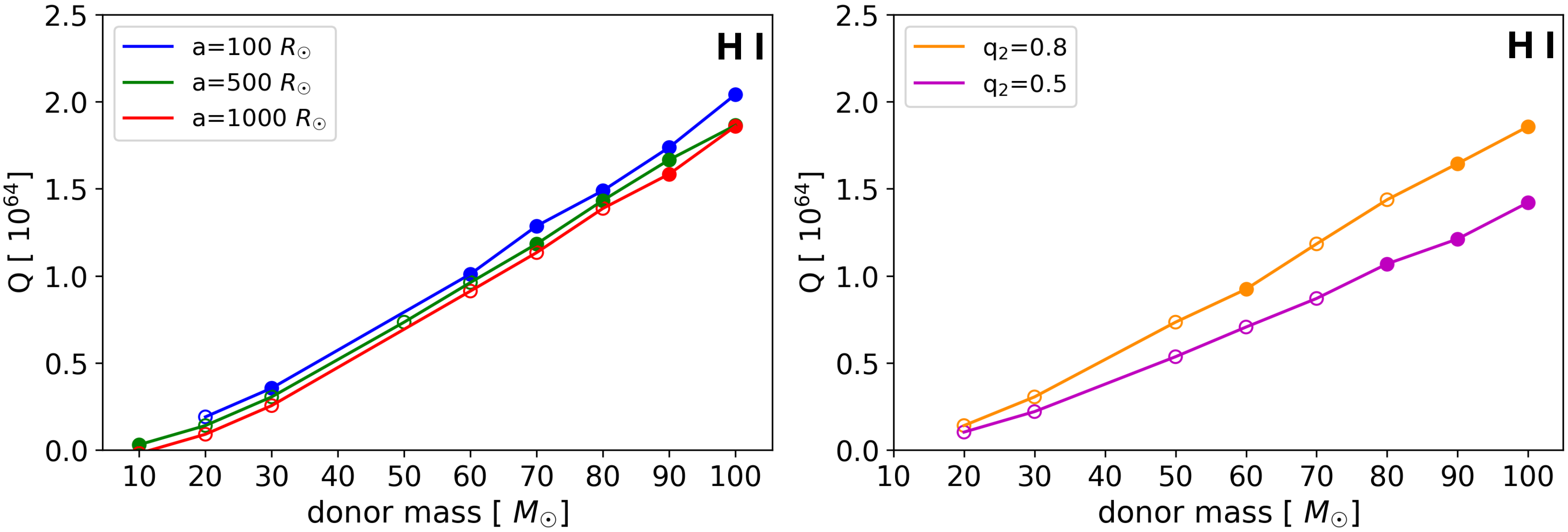}
\label{fig:HI}
\caption{Total HI ionizing photon yields for 0.001 \Zs\ binaries as a function of separation and metallicity. The filled and empty circles mark stable transfer and non-interacting binaries, respectively. As shown in the panel on the right, binary stars with larger $q_2$ have more massive companion stars with stronger ionizing fluxes and therefore greater photon yields.}
\end{figure*}

We show transfer rates for nine binaries with stable mass exchange in Figure~\ref{fig:transfer}.  In the left panel, those with smaller initial separations exchange mass sooner because the donor star fills its smaller Roche lobe at earlier times.  Binaries with higher metallicities also in general transfer mass sooner because their higher internal opacities lead to larger stellar radii \citep{ou2022critical}.  In the center panel, the 0.001 \Zs\ binary begins mass transfer before the 0.01 \Zs\ binary. This outlier occurs because the 0.001 \Zs\ donor star happens to fill its Roche lobe and activate the Ritter scheme before the 0.01 \Zs\ star does.  Binaries with smaller masses have longer main sequence lifetimes so mass transfer begins at later times. Transfer rates can vary from an average of 1.4 $\times$ 10$^{-4}$ \Ms\ yr$^{-1}$ to a maximum of 3.3 $\times$ 10$^{-3}$ \Ms\ yr$^{-1}$ in the $a = $ 100 $R_{\sun}$ binary in the left panel.

A number of the models exhibit sudden, large jumps in transfer rate by up to 4 - 6 orders in magnitude at the end of the life of the donor.  They are due to the rapid expansion of the envelope of the star prior to collapse or explosion.  From Equation~\ref{eq:xfer} it can be seen that the rate is an exponential function of the radius of the donor and can thus dramatically change with expansion at later times.  All the models exhibit an initial jump in transfer rate of 2 - 3 orders of magnitude at the onset of mass exchange because the expansion of the donor upon exiting the main sequence briefly exceeds its Roche lobe.

\subsection{Stellar Radii / Separations}

We show the evolution of the radii of the stars in a binary with an 80 \Ms\ donor, $q_2 =$ 0.8, $Z =$ 10$^{-2}$ \Zs, and $a =$ 800 \Rs\ in the upper panel of Figure~\ref{fig:radius}.  The donor star increases by a factor of ten in radius at 3.2 Myr, when it begins to burn helium.  It overflows its Roche lobe and begins to transfer mass to its companion until its core collapses after $\sim 0.3$ Myr.  At the onset of transfer, the radius of the companion star increases by about a factor of about five but then rapidly falls to a fration of its original value.  This happens because when the companion gains mass from the donor \texttt{MESA} adds it to its surface, which is in non-thermal equilibrium. The star then contracts on a timescale of about 1000 years, which is consistent with the Kelvin-Helmholtz time of the star, about 5000 years.  As shown in the lower panel of Figure~\ref{fig:radius}, at 3.25 Myr the separation of the stars in the binary evolves over a short period of time, 2500 yr.  Over this time the donor falls from 80 \Ms\ to 55 \Ms\ as it transfers mass to its companion at a rate of about 0.01 \Ms\ yr$^{-1}$. Thus, while mass transfer only occurs for a small fraction of the age of the donor, 31\% of its mass is transferred to its companion.   

\subsection{Ionizing Photon Yields}

We show total numbers of H~I, He~I, and He~II ionizing photons and LW photons for binaries of all masses with initial separations of 1000 \Rs\ and $q_{2} =$ 0.8 up to the death of the donor star in Figure~\ref{fig:ioniz}. They increase with total mass because of the higher temperatures of the more massive stars. In general, ionizing photon yields for a given mass are higher at lower metallicities because the stars are hotter and more compact. In contrast, LW yields increase with metallicity at a given mass because the stars are cooler and more of their photons lie in the $11.18 - 13.6$ eV band.  We compare H~I ionizing photon yields for the 0.001 \Zs\ binaries over a range of separations and mass ratios up to the death of the donor star in Figure~\ref{fig:HI}. They increase with both donor mass and mass ratio,  $q_{2}$.  Stars with larger masses have higher effective temperatures and larger surface areas, and thus greater photon yields. In a binary system, $q_2$ determines the mass of the companion star.  A binary with a larger $q_{2}$ will have a more massive companion than one with a lower $q_2$, and therefore stronger ionizing fluxes and greater photon yields. The yields rise as mass ratios increase because the more massive star can produce more ionizing UV flux.  Ionizing photon yields for our binaries are tabulated in the Appendix.


\begin{figure*}
\centering
\includegraphics[width=0.9\textwidth]{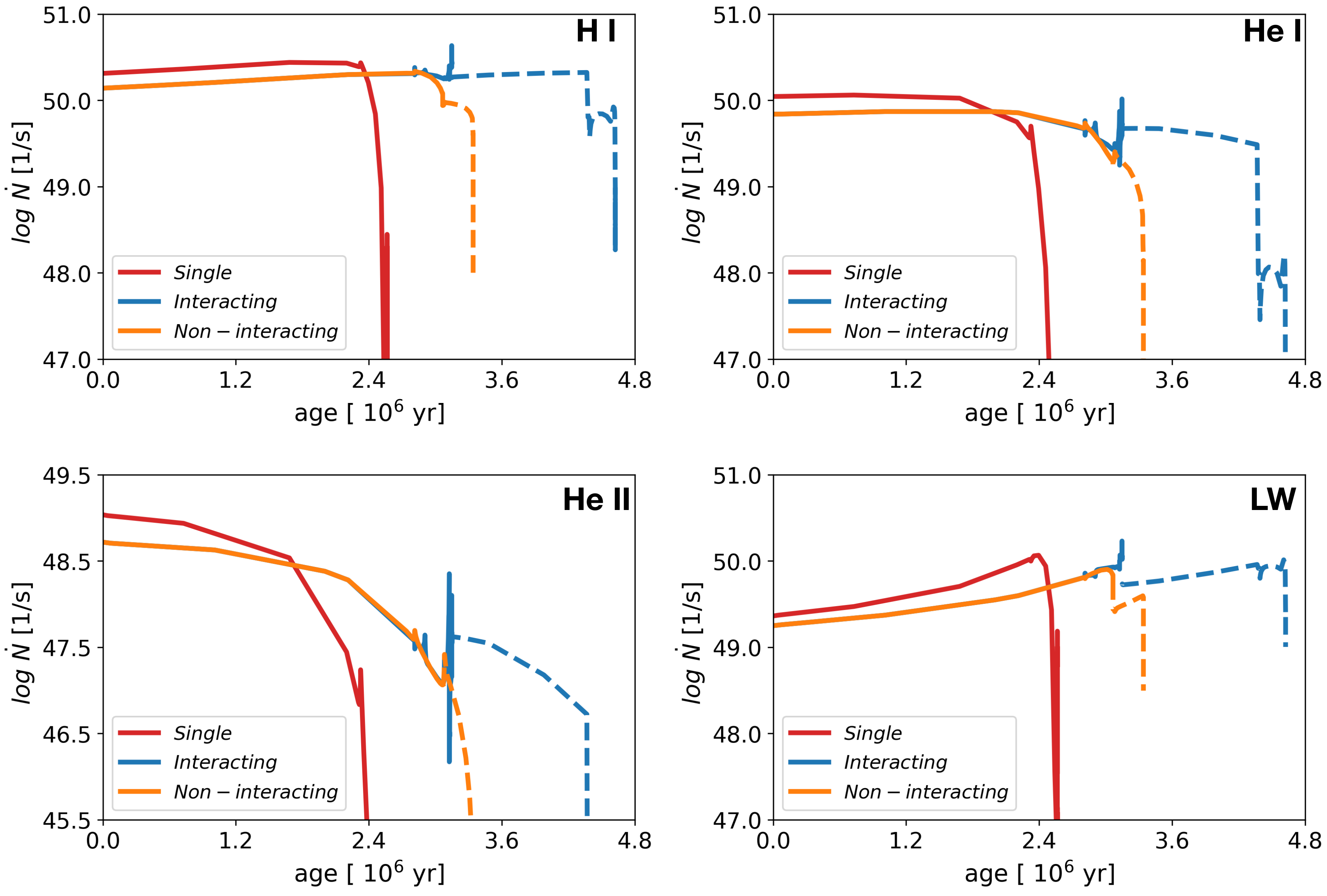}
\label{fig:single_binary}
\caption{UV emission rates for 162 \Ms\ zero-metallicity interacting and non-interacting binaries and a single star with the same total mass.  Here, $q_{2} =$ 0.8, and $a =$ 100 \Rs\ and 900 \Rs\ for the interacting and non-interacting binaries, respectively. The dashed lines are rates for the companion after it exits the main sequence. Mass transfer extends the life of the companion, so it produces more UV photons.}
\end{figure*}

\subsection{Ionizing Photon Rates}

Figure~\ref{fig:HI} shows that total ionizing photon yields for 0.001 \Zs\ binary stars decrease as the separation increases from 100 - 1000 \Rs, suggesting that mass transfer has some impact on them. To understand why this is so, we plot UV emission rates for 162 \Ms\ Pop~III non-interacting and stably interacting binaries to those of a single star with the same total mass in Figure~\ref{fig:single_binary} and compare their total photon yields in Table~\ref{tbl:2}.  In the table, "\textbf{S}" is the photon yield of the single star, "\textbf{I}" and "\textbf{NI}" are the yields of the binaries up to the collapse of the donor star, and "\textbf{I*}" and "\textbf{NI*}" are the final yields for the binaries at the collapse of the companion star.  Single stars have higher temperatures than those in binaries with the same total mass and thus have higher ionizing emission rates on the main sequence.  However, they also have shorter lifetimes and exit the main sequence before either star in the binary, after which, as shown in Figure~\ref{fig:single_binary}, their ionizing UV rates fall.  The net result is that single stars have somewhat higher total yields than binaries of equal mass up to the death of the donor star, but typically by no more than 50\%. 

As expected, the UV fluxes for interacting and non-interacting binary stars are similar up to the onset of mass transfer.  Mass exchange then alters the structures of both stars, typically resulting in pulsations in the donor that produce the luminosity waves visible in Figure~\ref{fig:single_binary} just before the star dies.  This process is short-lived, and the pulsations do not have much impact on ionizing photon yields, as seen in the \textbf{I} and \textbf{NI} rows in Table~\ref{tbl:2}.  However, even though it is brief, mass transfer extends the main sequence lifetime of the companion by providing it with additional fuel from the donor, which can add up to 50\% to the final yields of interacting binaries compared to those of non-interacting binaries of equal mass, as shown in the \textbf{I*} and \textbf{NI*} rows in Table~\ref{tbl:2}.  We note that while interacting binaries produce more H and He~I ionizing photons than single stars of equal mass, single stars produce more He~II ionizing photons because of their harder spectra.

\begin{figure*}
\centering
\includegraphics[width=\textwidth]{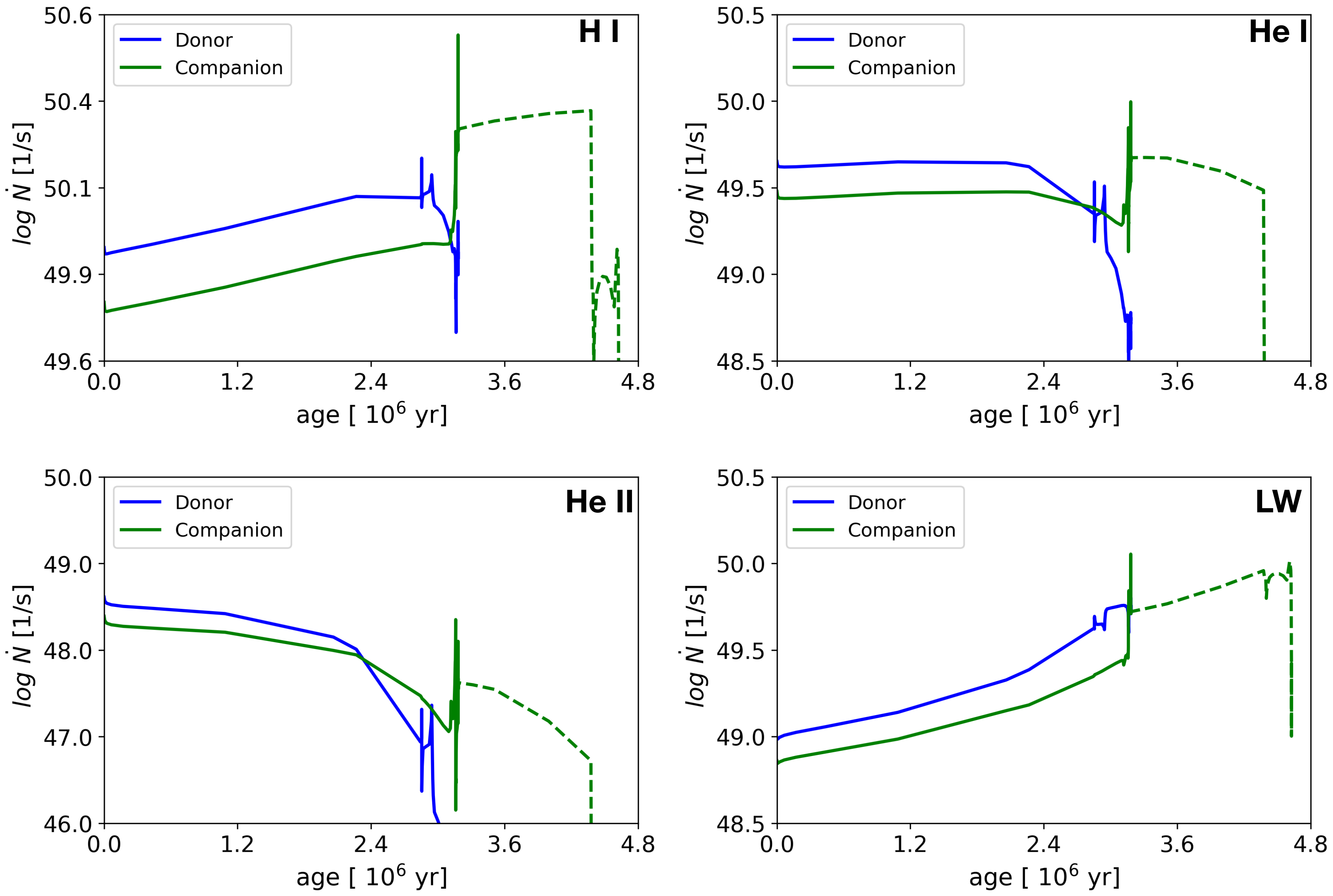}
\label{fig:donor_receptor}
\caption{UV emission rates from the donor and companion stars in a 90 \Ms\ zero-metallicity interacting binary with $q_2 =$ 0.8 and $a =$ 100 \Rs. After mass transfer, the companion star expands and increases in surface area to the point that its UV emission rates can exceed those of the donor star.}
\end{figure*}

We compare UV photon emission rates for donor and companion stars in 90 \Ms\ zero-metallicity interacting binary with $q_{2} =$ 0.8 and $a =$ 100 \Rs\ in Figure~\ref{fig:donor_receptor} and their total photon yields in Table~\ref{tbl:3}.  Initially the donor star, being more massive than its companion, has a higher effective temperature and larger surface area so it dominates the photon emission rates of the binary.  But when it exits the main sequence and expands in radius its effective temperature falls, which reduces its emission rates.  Its radius is then constrained by its Roche lobe during mass exchange.  The radius of the companion grows because of mass from the donor and its UV emission rates rise by up to an order in magnitude.

At the end of mass transfer the radius of the companion continues to increase, so its H~I, He~I, and LW photon emission rates keep rising.  However, its temperature does not increase much so it does not produce many He~II ionizing photons. In the absence of further accretion, the companion star evolves like an isolated single star and its UV emission rates fall as it enters the post main sequence.  Table 3 shows that while the donor star has higher UV photon yields over its lifetime, the companion star dominates the total UV yields of the binary because mass transfer extended its main sequence lifetime.  As the two stars evolve over their main sequence lifetimes, their radii and surface temperatures gradually rise and fall, respectively.  The increase in the H~I and LW photon emission rates over this interval in Figures~\ref{fig:single_binary} and \ref{fig:donor_receptor} are due to their increase in surface area while the stability or fall of the He~I and He~II emission rates are due to the decrease in surface temperature. 


\begin{table}
\begin{center}
\begin{tabular}{ |c|c|c|c|c| } 
\hline
& H~I & He~I & He~II & LW \\ 
\hline
\textbf{S} & $1.91 \times 10^{64}$ & $7.65 \times 10^{63}$ & $4.57 \times 10^{62}$  & $3.90 \times 10^{63}$\\ 
\textbf{I} & $1.75 \times 10^{64}$ & $6.65 \times 10^{63}$ & $3.02 \times 10^{62}$  & $3.64 \times 10^{63}$\\ 
\textbf{NI} & $1.69 \times 10^{64}$ & $6.55 \times 10^{63}$ & $3.01 \times 10^{62}$  & $3.37 \times 10^{63}$\\ 
\textbf{I*} & $2.56 \times 10^{64}$ & $8.22 \times 10^{63}$ & $3.11 \times 10^{62}$  & $6.87 \times 10^{63}$\\ 
\textbf{NI*} & $1.77 \times 10^{64}$ & $6.68 \times 10^{63}$ & $3.01 \times 10^{62}$  & $3.65 \times 10^{63}$\\ 
\hline
\end{tabular}
\end{center}
\caption{UV photon yields for 162 \Ms\ Pop~III interacting and non-interacting binaries and for a single 162 \Ms\ star.  Here, $q_2 =$ 0.8 and $a =$ 100 \Rs\ and 900 \Rs\ for the interacting and non-interacting systems, respectively.  "\textbf{S}" is the photon yield of the single star, "\textbf{I}" and "\textbf{NI}" are the yields of the binaries up to the collapse of the donor star, and "\textbf{I*}" and "\textbf{NI*}" are the final yields for the binaries at the time of collapse of the companion star.}
\label{tbl:2}
\end{table}


We fit the ionizing and LW photon yields up to the death of the donor star, N, with a simple power-law function of donor mass, $M^{\alpha}$, in Figure \ref{fig:fitting}.  Fits to the H~I and LW photon yields are limited to 10 - 80 \Ms\ because the power law overestimates the yields at higher masses.  Also, the high temperature thresholds for He~I and He~II photon production mean that stars below 10 \Ms\ cannot efficiently emit these photons, so we constrain the range of validity to of these fits to 20 - 80 \Ms. The power law index increases with photon energy and metallicity. We compare photon yields for binaries with single stars of the same total mass in Figure \ref{fig:fitting_single}. Massive single stars can produce more ionizing photons than their corresponding binaries, as shown by their higher power-law index in all models. In Figure 11, we also show the power-law fits as a function of stellar mass. The mass covered by our analysis is 10 - 150 \Ms\ for H I and LW photon yields and 30 - 150 \Ms\ for He I and He II yields. Notably, the $q_2 = 0.8$ and 0.5 binaries exhibit similar photon yields. However, for He II photons, the $q_2 = 0.5$ binaries have higher yields because binaries with a lower mass ratio have a more massive donor star that produces a greater number of He II photons. These fits can be used in cosmological simulations with radiative feedback at high redshifts.

We note that these fits underestimate the true photon yields of the binaries to some degree because they exclude contributions by the companions over their post-donor lifetimes, which, as shown in Figures~\ref{fig:single_binary}, \ref{fig:donor_receptor} and Table~\ref{tbl:2} can enhance the yields by up to 100\%. They also ignore ionizing UV (and X-rays) from the black hole of the donor, which are difficult to quantify.  Nevertheless, they capture the significant enhancement in yields due to the rejuvenation of the companion star by the donor over its lifetime.

\subsection{Effects of Mass Transfer on Stellar Evolution}

\begin{figure*}
\centering
\includegraphics[width=\textwidth]{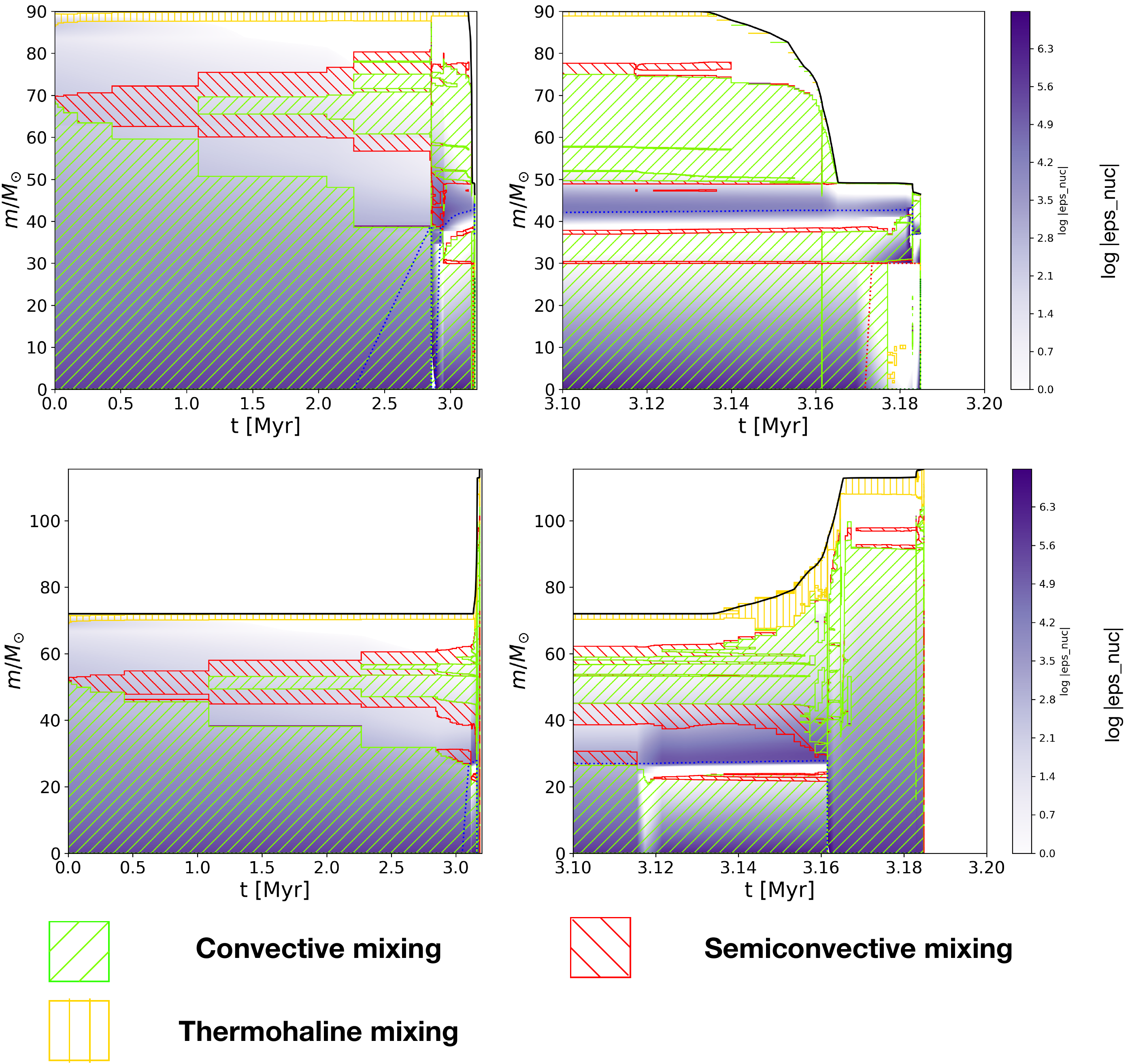}
\label{fig:kipp}
\caption{Kippenhahn diagrams for the stars in the zero-metallicity 90 \Ms\ binary with $q_2 = $ 0.8 and initial separation $a =$ 100 \Rs. The upper-left panel shows the evolution of the donor star and the upper-right zooms in the period of mass transfer.  The bottom two panels show the same stages of evolution for the companion star.}
\end{figure*}

\begin{figure*}
\centering
\includegraphics[width=\textwidth]{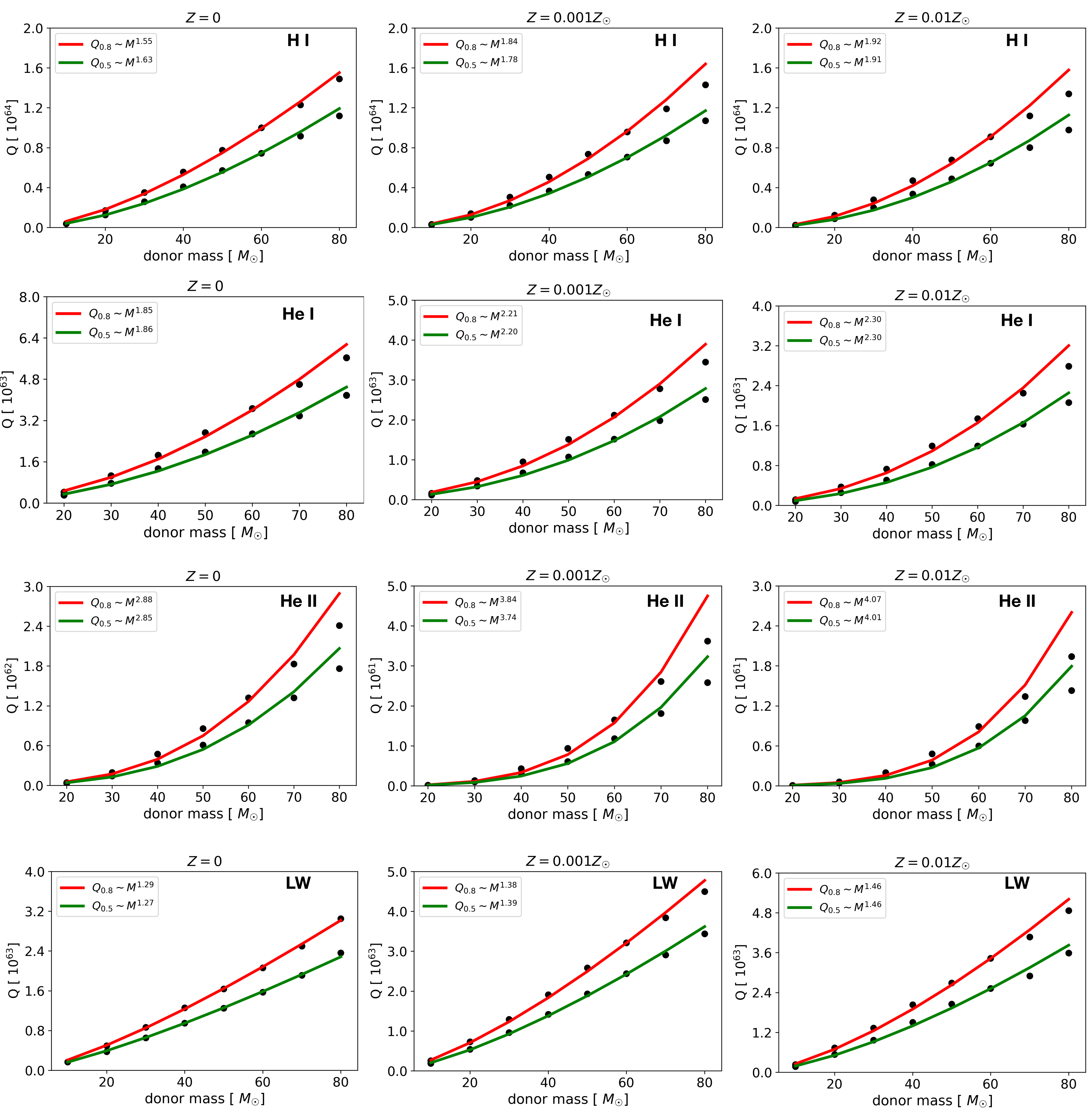}
\label{fig:fitting}
\caption{Fits to the binary photon yields up to the death of the donor star. The $Q_{0.8}$ and $Q_{0.5}$ lines are the fits ($Q \sim M^{\alpha}$) for the $q_2 =$ 0.8 and 0.5 binaries, respectively.}
\end{figure*}

\begin{figure*}
\centering
\includegraphics[width=\textwidth]{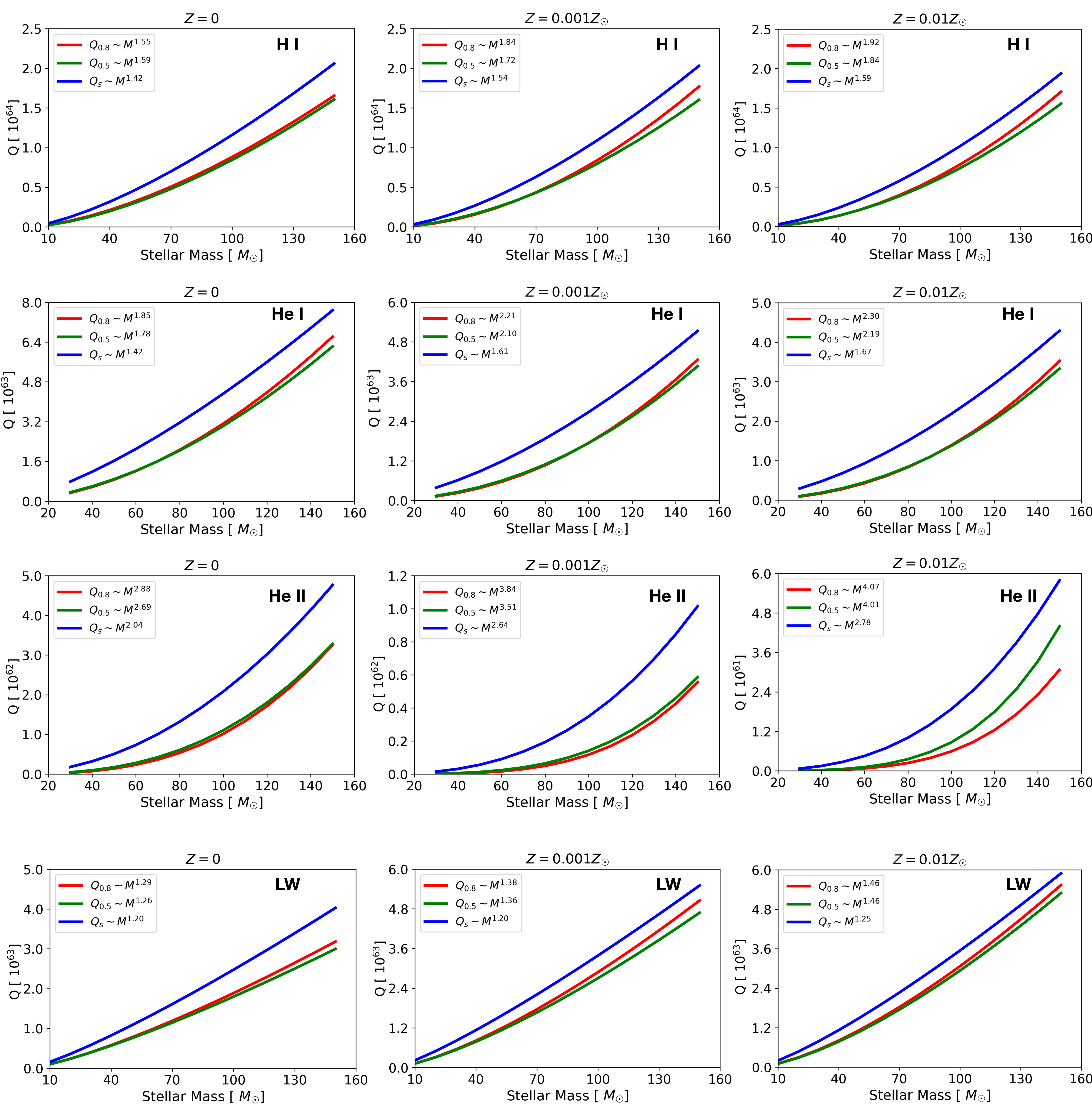}
\label{fig:fitting_single}
\caption{Ionizing and LW UV photon yields for binaries and single stars of equal mass. As in Figure \ref{fig:fitting}, $Q_{s}$ is the power-law fit ($Q \sim M^{\alpha}$) for single stars. The $\alpha$ for binaries is smaller than that of single stars of equal mass.}
\end{figure*}

To understand how mass transfer affects stellar evolution in binaries, we show Kippenhahn diagrams for a zero-metallicity 90 \Ms\ binary with $q_2 = $ 0.8 and initial separation $a =$ 100 \Rs\ in Figure~\ref{fig:kipp}.  When the donor star depletes its central hydrogen, its core contracts because of gravity, which triggers central helium burning and the exit of the star from the main sequence.  Hydrogen shell burning also begins and causes the radius of the star to grow by a factor of 100, so it becomes a red supergiant.  At the onset of mass transfer, the temperature of the donor falls because mass loss reduces the gravity of the star, and in turn the energy production rates in the shell.  Nuclear burning continues in the core of the donor until iron forms. The core itself is detached from mass transfer.  

During mass exchange, the donor transfers most of its hydrogen envelope to its companion and becomes hydrogen-poor.  The loss of the envelope could produce a Type Ib or Ic supernova.  At the same time, the companion star remains on the main sequence. Mass transfer increases its radius and forms an outer semiconvective zone, which affects hydrogen shell burning and the mass of helium core. This in turn alters the evolutionary track of the donor star toward a yellow supergiant. Since the final mass of this star is $\sim$ 115 \Ms, it likely dies as a black hole without exploding.


\begin{table}[h]
\begin{center}
\vspace{0.2in}
\begin{tabular}{ |c|c|c|c|c| } 
\hline
& H~I & He~I & He~II & LW \\ 
\hline
D & $1.03 \times 10^{64}$ & $3.84 \times 10^{63}$ & $1.82 \times 10^{62}$  & $2.25 \times 10^{63}$\\ 
C & $7.26 \times 10^{63}$ & $2.81 \times 10^{63}$ & $1.19 \times 10^{62}$  & $1.39 \times 10^{63}$\\
C* & $1.53 \times 10^{64}$ & $4.38 \times 10^{63}$ & $1.28 \times 10^{62}$  & $4.62 \times 10^{63}$\\
\hline
\end{tabular}
\end{center}
\vspace{0.1in}
\caption{Contributions of the donor (D) and companion (C) stars to the total photon yields of the interacting binary. C* includes the contribution from the companion star after the collapse of the donor.}
\label{tbl:3}
\end{table}


\section{Discussion}

Besides mass transfer, binary stars can in principle gain mass from each other by stellar winds.  However, the winds are mainly line driven, with mass loss rates that are proportional to the metallicity of the star \citep{vink01,muij12}.  At $Z \lesssim$ 10$^{-2}$ \Zs, mass loss rates are at most $\sim$ 10$^{-6}$ \Ms\ yr$^{-1}$, far lower than typical Roche lobe overflow rates of 10$^{-4}$ - 10$^{-3}$ \Ms\ yr$^{-1}$, so we neglect them in our models.

Algol-type binaries contain a main sequence star as the primary and a post-main sequence star that has filled its Roche lobe as the secondary, with both stars in an accretion disk.  A number of Algol-type eclipsing binaries have been found  \citep{zola1992ku,ghoreyshi2011accretion,atwood2012modeling}. These studies used observational photometric data of their accretion disks to determine that these objects were low-mass binaries. Our massive binaries therefore cannot be considered to be Algol-type binaries.

\citet{chen2015cosmological} considered the impact of Pop~III binaries on primordial structure formation in the early Universe.  They used stellar evolution models from \citet{heger2010nucleosynthesis} to estimate ionizing UV feedback due to 45 \Ms\ $+$ 15 \Ms\ and 30 \Ms\ $+$ 30 \Ms\ Pop III binary systems.  However, they treated their binaries as two isolated stars without accounting for the effect of mass transfer on their ionizing UV photon budgets.  Our study shows that they therefore likely underestimated total ionizing UV from binaries by up to 50\%, which resulted in smaller H~II regions and higher ambient densities for supernovae, so metals from explosions would not have propagated as far into the early IGM in their models.  Furthermore, we find that mass exchange can nearly double total LW photon yields, so Pop III and EMP binaries may have been particularly effective at suppressing the formation of new Pop~III stars in their vicinity.  UV photon yields from Figure~\ref{fig:ioniz} can be used to implement recipes for radiative feedback due to Pop~III binaries in future cosmological simulations.

In calculating UV photon yields we treated our stars as black bodies, but non-LTE radiative transfer models have shown that absorption by stellar atmospheres can affect these fluxes \citep{s02}.  However, such calculations would have to be done on the fly to capture ionizing UV escape fractions during mass transfer because the atmosphere of the companion star is constantly changing, so this approach is too costly in the near term.  We also adopted a simplified prescription for mass transfer that in reality can require detailed modeling of accretion disks that may be highly turbulent and have magnetic fields, which is beyond the scope of our study.  Many of our interacting binaries enter common envelope evolution, especially those with high metallicities, but this phase cannot currently be modeled with \texttt{MESA}.  Sophisticated 2D or 3D hydrodynamics simulations are required to study stellar overlap.

\section{Conclusion}

We find that interactions ranging from stable mass transfer to common envelope evolution can occur in Pop~III and EMP binary stars in the early Universe.  In particular, mass exchange in these systems can enhance their total H~I, He~I, He~II, and LW photon yields by up to $46\%$, $24\%$, $2\%$, and $89\%$, respectively, because of the extension of the life of the companion star.  Mass transfer in Pop~III and EMP binaries can therefore enhance early cosmological reionization and suppress the formation of later generations of primordial stars.  Mass exchange can also affect their nucleosynthetic yields by changing the masses of the stars at death.  Pop~III stars from 8 - 25 \Ms\ die as Type II core-collapse SNe and 90 - 260 \Ms\ stars explode as pair-instability SNe \citep{hw02,wet12a, wet12c,wet12b}.   If mass transfer between the stars promotes or removes them into or out of these mass ranges, it could drastically alter metal yields due to early binaries.  The effect of binary interactions on chemical yields from early populations of stars will be investigated in future studies.

\acknowledgments

We thank Yi Chou and You-Hua Chu for useful comments and suggestions. This research is supported by the National Science and Technology Council, Taiwan under grant number MOST 110-2112-M-001-068-MY3 and by the Academia Sinica, Taiwan under the career development award AS-CDA-111-M04.  Our simulations were performed at the National Energy Research Scientific Computing Center (NERSC), a U.S. Department of Energy Office of Science User Facility operated under Contract No. DE-AC02-05CH11231, and on the TIARA Cluster at the Academia Sinica Institute of Astronomy and Astrophysics (ASIAA).


\bibliographystyle{yahapj}
\bibliography{refs}

\appendix

Here we show twelve tables of ionizing (H I, He I, He II) and LW UV photon yields for our binary star models for $Z =$ 0, 0.1, 0.01 \Zs. These tables exclude common envelope candidates. The yields for stably-interacting models are calculated from the formation of the binary star until the donor star reaches carbon burning or iron-core collapse. We also provide photon yields for 10 to 180 \Ms\ single stars in Table~\ref{tbl:16}

\begin{center}
\begin{longtable}{ |c|c|c|c|c|c| }
	\hline
	\multicolumn{6}{|c|}{Z = 0, H I} \\
	\hline $M_{1} [\Ms]$& $M_{2} [\Ms]$ &a [$R_{\sun}$]&$Q_{t}$ &$Q_{d}$ &$Q_{c}$ \\ \hline
	\endfirsthead
	
	\multicolumn{6}{c}%
	{{\bfseries \tablename\ \thetable{} -- continued from previous page}} \\
	\hline \multicolumn{1}{|c|}{$M_{1} [\Ms]$} & \multicolumn{1}{c|}{$M_{2} [\Ms]$} & \multicolumn{1}{c|}{a [$R_{\sun}$]} & \multicolumn{1}{c|}{$Q_{t}$ } & \multicolumn{1}{c|}{$Q_{d}$ } & \multicolumn{1}{c|}{$Q_{c}$ }\\ \hline 
	\endhead
	
	\hline \multicolumn{6}{|r|}{{Continued on next page}} \\ \hline
	\endfoot
	
	\hline \hline
	\endlastfoot
	
	10&5&900&$3.75\times10^{62}$&$3.62\times10^{62}$&$1.28\times10^{61}$\\
        10&5&1000&$3.75\times10^{62}$&$3.62\times10^{62}$&$1.28\times10^{61}$\\
	\hline
	20&16&100&$1.69\times10^{63}$&$1.13\times10^{63}$&$5.62\times10^{62}$\\
        20&16&200&$1.69\times10^{63}$&$1.13\times10^{63}$&$5.62\times10^{62}$\\
        20&16&300&$1.69\times10^{63}$&$1.13\times10^{63}$&$5.62\times10^{62}$\\
        20&16&400&$1.69\times10^{63}$&$1.13\times10^{63}$&$5.62\times10^{62}$\\
        20&16&500&$1.69\times10^{63}$&$1.13\times10^{63}$&$5.62\times10^{62}$\\
        20&16&600&$1.69\times10^{63}$&$1.13\times10^{63}$&$5.62\times10^{62}$\\
\end{longtable}
\end{center}

\begin{center}
\begin{longtable}{ |c|c|c|c|c|c| }
	\hline
	\multicolumn{6}{|c|}{Z = 0, He I} \\
	\hline $M_{1} [\Ms]$& $M_{2} [\Ms]$ &a [$R_{\sun}$]&$Q_{t}$ &$Q_{d}$ &$Q_{c}$ \\ \hline
	\endfirsthead
	
	\multicolumn{6}{c}%
	{{\bfseries \tablename\ \thetable{} -- continued from previous page}} \\
	\hline \multicolumn{1}{|c|}{$M_{1} [\Ms]$} & \multicolumn{1}{c|}{$M_{2} [\Ms]$} & \multicolumn{1}{c|}{a [$R_{\sun}$]} & \multicolumn{1}{c|}{$Q_{t}$ } & \multicolumn{1}{c|}{$Q_{d}$ } & \multicolumn{1}{c|}{$Q_{c}$ }\\ \hline 
	\endhead
	
	\hline \multicolumn{6}{|r|}{{Continued on next page}} \\ \hline
	\endfoot
	
	\hline \hline
	\endlastfoot
	
	10&5&900&$5.35\times10^{61}$&$5.31\times10^{61}$&$4.12\times10^{59}$\\
        10&5&1000&$5.35\times10^{61}$&$5.31\times10^{61}$&$4.12\times10^{59}$\\
	\hline
	20&16&100&$4.21\times10^{62}$&$2.88\times10^{62}$&$1.33\times10^{62}$\\
        20&16&200&$4.21\times10^{62}$&$2.88\times10^{62}$&$1.33\times10^{62}$\\
        20&16&300&$4.21\times10^{62}$&$2.88\times10^{62}$&$1.33\times10^{62}$\\
        20&16&400&$4.21\times10^{62}$&$2.88\times10^{62}$&$1.33\times10^{62}$\\
        20&16&500&$4.21\times10^{62}$&$2.88\times10^{62}$&$1.33\times10^{62}$\\
        20&16&600&$4.21\times10^{62}$&$2.88\times10^{62}$&$1.33\times10^{62}$\\
\end{longtable}
\end{center}

\begin{center}
\begin{longtable}{ |c|c|c|c|c|c| }
	\hline
	\multicolumn{6}{|c|}{Z = 0, He II} \\
	\hline $M_{1} [\Ms]$& $M_{2} [\Ms]$ &a [$R_{\sun}$]&$Q_{t}$ &$Q_{d}$ &$Q_{c}$ \\ \hline
	\endfirsthead
	
	\multicolumn{6}{c}%
	{{\bfseries \tablename\ \thetable{} -- continued from previous page}} \\
	\hline \multicolumn{1}{|c|}{$M_{1} [\Ms]$} & \multicolumn{1}{c|}{$M_{2} [\Ms]$} & \multicolumn{1}{c|}{a [$R_{\sun}$]} & \multicolumn{1}{c|}{$Q_{t}$ } & \multicolumn{1}{c|}{$Q_{d}$ } & \multicolumn{1}{c|}{$Q_{c}$ }\\ \hline 
	\endhead
	
	\hline \multicolumn{6}{|r|}{{Continued on next page}} \\ \hline
	\endfoot
	
	\hline \hline
	\endlastfoot
	
	10&5&900&$1.01\times10^{59}$&$1.01\times10^{59}$&$1.29\times10^{55}$\\
        10&5&1000&$1.01\times10^{59}$&$1.01\times10^{59}$&$1.24\times10^{55}$\\
	\hline
	20&16&100&$4.26\times10^{60}$&$3.19\times10^{60}$&$1.07\times10^{60}$\\
        20&16&200&$4.26\times10^{60}$&$3.19\times10^{60}$&$1.07\times10^{60}$\\
        20&16&300&$4.26\times10^{60}$&$3.19\times10^{60}$&$1.07\times10^{60}$\\
        20&16&400&$4.26\times10^{60}$&$3.19\times10^{60}$&$1.07\times10^{60}$\\
        20&16&500&$4.26\times10^{60}$&$3.19\times10^{60}$&$1.07\times10^{60}$\\
        20&16&600&$4.26\times10^{60}$&$3.19\times10^{60}$&$1.07\times10^{60}$\\
\end{longtable}
\end{center}

\begin{center}
\begin{longtable}{ |c|c|c|c|c|c| }
	\hline
	\multicolumn{6}{|c|}{Z = 0, LW} \\
	\hline $M_{1} [\Ms]$& $M_{2} [\Ms]$ &a [$R_{\sun}$]&$Q_{t}$ &$Q_{d}$ &$Q_{c}$ \\ \hline
	\endfirsthead
	
	\multicolumn{6}{c}%
	{{\bfseries \tablename\ \thetable{} -- continued from previous page}} \\
	\hline \multicolumn{1}{|c|}{$M_{1} [\Ms]$} & \multicolumn{1}{c|}{$M_{2} [\Ms]$} & \multicolumn{1}{c|}{a [$R_{\sun}$]} & \multicolumn{1}{c|}{$Q_{t}$ } & \multicolumn{1}{c|}{$Q_{d}$ } & \multicolumn{1}{c|}{$Q_{c}$ }\\ \hline 
	\endhead
	
	\hline \multicolumn{6}{|r|}{{Continued on next page}} \\ \hline
	\endfoot
	
	\hline \hline
	\endlastfoot
	
	10&5&900&$1.69\times10^{62}$&$1.57\times10^{62}$&$1.22\times10^{61}$\\
        10&5&1000&$1.69\times10^{62}$&$1.57\times10^{62}$&$1.22\times10^{61}$\\
	\hline
	20&16&100&$4.96\times10^{62}$&$3.26\times10^{62}$&$1.70\times10^{62}$\\
        20&16&200&$4.96\times10^{62}$&$3.26\times10^{62}$&$1.70\times10^{62}$\\
        20&16&300&$4.96\times10^{62}$&$3.26\times10^{62}$&$1.70\times10^{62}$\\
        20&16&400&$4.96\times10^{62}$&$3.26\times10^{62}$&$1.70\times10^{62}$\\
        20&16&500&$4.96\times10^{62}$&$3.26\times10^{62}$&$1.70\times10^{62}$\\
        20&16&600&$4.96\times10^{62}$&$3.26\times10^{62}$&$1.70\times10^{62}$\\
\end{longtable}
\end{center}

\begin{center}
\begin{longtable}{ |c|c|c|c|c|c| }
	\hline
	\multicolumn{6}{|c|}{Z = 0.001 \Zs, H I} \\
	\hline $M_{1} [\Ms]$& $M_{2} [\Ms]$ &a [$R_{\sun}$]&$Q_{t}$ &$Q_{d}$ &$Q_{c}$ \\ \hline
	\endfirsthead
		
	\multicolumn{6}{c}%
	{{\bfseries \tablename\ \thetable{} -- continued from previous page}} \\
	\hline \multicolumn{1}{|c|}{$M_{1} [\Ms]$} & \multicolumn{1}{c|}{$M_{2} [\Ms]$} & \multicolumn{1}{c|}{a [$R_{\sun}$]} & \multicolumn{1}{c|}{$Q_{t}$ } & \multicolumn{1}{c|}{$Q_{d}$ } & \multicolumn{1}{c|}{$Q_{c}$ }\\ \hline 
	\endhead
		
	\hline \multicolumn{6}{|r|}{{Continued on next page}} \\ \hline
	\endfoot
		
	\hline \hline
	\endlastfoot
		
	10&8&500&$3.01\times10^{62}$&$2.16\times10^{62}$&$8.46\times10^{61}$\\
        10&8&900&$3.01\times10^{62}$&$2.16\times10^{62}$&$8.48\times10^{61}$\\
        10&8&1000&$3.01\times10^{62}$&$2.16\times10^{62}$&$8.48\times10^{61}$\\
	\hline
	10&5&900&$2.25\times10^{62}$&$2.16\times10^{62}$&$8.86\times10^{60}$\\
        10&5&1000&$2.25\times10^{62}$&$2.16\times10^{62}$&$8.86\times10^{60}$\\
	\hline
	20&16&100&$1.40\times10^{63}$&$9.47\times10^{62}$&$4.52\times10^{62}$\\
        20&16&200&$1.40\times10^{63}$&$9.49\times10^{62}$&$4.52\times10^{62}$\\
        20&16&300&$1.40\times10^{63}$&$9.49\times10^{62}$&$4.52\times10^{62}$\\
\end{longtable}
\end{center}

\begin{center}
\begin{longtable}{ |c|c|c|c|c|c| }
	\hline
	\multicolumn{6}{|c|}{Z = 0.001 \Zs, He I} \\
	\hline $M_{1} [\Ms]$& $M_{2} [\Ms]$ &a [$R_{\sun}$]&$Q_{t}$ &$Q_{d}$ &$Q_{c}$ \\ \hline
	\endfirsthead
		
	\multicolumn{6}{c}%
	{{\bfseries \tablename\ \thetable{} -- continued from previous page}} \\
	\hline \multicolumn{1}{|c|}{$M_{1} [\Ms]$} & \multicolumn{1}{c|}{$M_{2} [\Ms]$} & \multicolumn{1}{c|}{a [$R_{\sun}$]} & \multicolumn{1}{c|}{$Q_{t}$ } & \multicolumn{1}{c|}{$Q_{d}$ } & \multicolumn{1}{c|}{$Q_{c}$ }\\ \hline 
	\endhead
		
	\hline \multicolumn{6}{|r|}{{Continued on next page}} \\ \hline
	\endfoot
		
	\hline \hline
	\endlastfoot
		
	10&8&500&$1.42\times10^{61}$&$1.08\times10^{61}$&$3.37\times10^{60}$\\ 
        10&8&900&$1.42\times10^{61}$&$1.08\times10^{61}$&$3.37\times10^{60}$\\
        10&8&1000&$1.42\times10^{61}$&$1.08\times10^{61}$&$3.37\times10^{60}$\\ 
	\hline
	10&5&900&$1.09\times10^{61}$&$1.08\times10^{61}$&$1.41\times10^{59}$\\
        10&5&1000&$1.09\times10^{61}$&$1.08\times10^{61}$&$1.41\times10^{59}$\\
	\hline
	20&16&100&$1.59\times10^{62}$&$1.12\times10^{62}$&$4.73\times10^{61}$\\
        20&16&200&$1.59\times10^{62}$&$1.12\times10^{62}$&$4.73\times10^{61}$\\
        20&16&300&$1.59\times10^{62}$&$1.12\times10^{62}$&$4.73\times10^{61}$\\
\end{longtable}
\end{center}

\begin{center}
\begin{longtable}{ |c|c|c|c|c|c| }
	\hline
	\multicolumn{6}{|c|}{Z = 0.001 \Zs, He II} \\
	\hline $M_{1} [\Ms]$& $M_{2} [\Ms]$ &a [$R_{\sun}$]&$Q_{t}$ &$Q_{d}$ &$Q_{c}$ \\ \hline
	\endfirsthead
		
	\multicolumn{6}{c}%
	{{\bfseries \tablename\ \thetable{} -- continued from previous page}} \\
	\hline \multicolumn{1}{|c|}{$M_{1} [\Ms]$} & \multicolumn{1}{c|}{$M_{2} [\Ms]$} & \multicolumn{1}{c|}{a [$R_{\sun}$]} & \multicolumn{1}{c|}{$Q_{t}$ } & \multicolumn{1}{c|}{$Q_{d}$ } & \multicolumn{1}{c|}{$Q_{c}$ }\\ \hline 
	\endhead
		
	\hline \multicolumn{6}{|r|}{{Continued on next page}} \\ \hline
	\endfoot
		
	\hline \hline
	\endlastfoot
		
	10&8&500&$1.32\times10^{57}$&$1.17\times10^{57}$&$1.52\times10^{56}$\\
        10&8&900&$1.32\times10^{57}$&$1.17\times10^{57}$&$1.51\times10^{56}$\\
        10&8&1000&$1.32\times10^{57}$&$1.17\times10^{57}$&$1.51\times10^{56}$\\
	\hline
	10&5&900&$1.17\times10^{57}$&$1.17\times10^{57}$&$4.80\times10^{53}$\\
        10&5&1000&$1.17\times10^{57}$&$1.17\times10^{57}$&$4.80\times10^{53}$\\
	\hline
	20&16&100&$1.67\times10^{59}$&$1.34\times10^{59}$&$3.28\times10^{58}$\\
        20&16&200&$1.67\times10^{59}$&$1.34\times10^{59}$&$3.28\times10^{58}$\\
        20&16&300&$1.67\times10^{59}$&$1.34\times10^{59}$&$3.28\times10^{58}$\\
\end{longtable}
\end{center}

\begin{center}
\begin{longtable}{ |c|c|c|c|c|c| }
	\hline
	\multicolumn{6}{|c|}{Z = 0.001 \Zs, LW} \\
	\hline $M_{1} [\Ms]$& $M_{2} [\Ms]$ &a [$R_{\sun}$]&$Q_{t}$ &$Q_{d}$ &$Q_{c}$ \\ \hline
	\endfirsthead
		
	\multicolumn{6}{c}%
	{{\bfseries \tablename\ \thetable{} -- continued from previous page}} \\
	\hline \multicolumn{1}{|c|}{$M_{1} [\Ms]$} & \multicolumn{1}{c|}{$M_{2} [\Ms]$} & \multicolumn{1}{c|}{a [$R_{\sun}$]} & \multicolumn{1}{c|}{$Q_{t}$ } & \multicolumn{1}{c|}{$Q_{d}$ } & \multicolumn{1}{c|}{$Q_{c}$ }\\ \hline 
	\endhead
		
	\hline \multicolumn{6}{|r|}{{Continued on next page}} \\ \hline
	\endfoot
		
	\hline \hline
	\endlastfoot
		
	10&8&500&$2.52\times10^{62}$&$1.78\times10^{62}$&$7.39\times10^{61}$\\
        10&8&900&$2.52\times10^{62}$&$1.78\times10^{62}$&$7.41\times10^{61}$\\
        10&8&1000&$2.52\times10^{62}$&$1.78\times10^{62}$&$7.41\times10^{61}$\\
	\hline
	10&5&900&$1.89\times10^{62}$&$1.78\times10^{62}$&$1.13\times10^{61}$\\
        10&5&1000&$1.89\times10^{62}$&$1.78\times10^{62}$&$1.13\times10^{61}$\\
	\hline
	20&16&100&$7.30\times10^{62}$&$4.89\times10^{62}$&$2.41\times10^{62}$\\
        20&16&200&$7.30\times10^{62}$&$4.89\times10^{62}$&$2.41\times10^{62}$\\
        20&16&300&$7.30\times10^{62}$&$4.89\times10^{62}$&$2.41\times10^{62}$\\
\end{longtable}
\end{center}

\begin{center}
\begin{longtable}{ |c|c|c|c|c|c| }
	\hline
	\multicolumn{6}{|c|}{Z = 0.01 \Zs, H I} \\
	\hline $M_{1} [\Ms]$& $M_{2} [\Ms]$ &a [$R_{\sun}$]&$Q_{t}$ &$Q_{d}$ &$Q_{c}$ \\ \hline
	\endfirsthead
		
	\multicolumn{6}{c}%
	{{\bfseries \tablename\ \thetable{} -- continued from previous page}} \\
	\hline \multicolumn{1}{|c|}{$M_{1} [\Ms]$} & \multicolumn{1}{c|}{$M_{2} [\Ms]$} & \multicolumn{1}{c|}{a [$R_{\sun}$]} & \multicolumn{1}{c|}{$Q_{t}$ } & \multicolumn{1}{c|}{$Q_{d}$ } & \multicolumn{1}{c|}{$Q_{c}$ }\\ \hline 
	\endhead
		
	\hline \multicolumn{6}{|r|}{{Continued on next page}} \\ \hline
	\endfoot
		
	\hline \hline
	\endlastfoot
		
	10&8&100&$2.40\times10^{62}$&$1.70\times10^{62}$&$7.05\times10^{61}$\\
	\hline
	10&5&900&$1.76\times10^{62}$&$1.70\times10^{62}$&$6.46\times10^{60}$\\
        10&5&1000&$1.76\times10^{62}$&$1.70\times10^{62}$&$6.46\times10^{60}$\\
	\hline
	20&16&100&$1.38\times10^{63}$&$8.61\times10^{62}$&$5.17\times10^{62}$\\
        20&16&200&$1.23\times10^{63}$&$8.27\times10^{62}$&$3.99\times10^{62}$\\
        20&16&300&$1.22\times10^{63}$&$8.21\times10^{62}$&$3.99\times10^{62}$\\
        20&16&400&$1.22\times10^{63}$&$8.21\times10^{62}$&$3.99\times10^{62}$\\
        20&16&500&$1.22\times10^{63}$&$8.21\times10^{62}$&$3.99\times10^{62}$\\
\end{longtable}
\end{center}

\begin{center}
\begin{longtable}{ |c|c|c|c|c|c| }
	\hline
	\multicolumn{6}{|c|}{Z = 0.01 \Zs, He I} \\
	\hline $M_{1} [\Ms]$& $M_{2} [\Ms]$ &a [$R_{\sun}$]&$Q_{t}$ &$Q_{d}$ &$Q_{c}$ \\ \hline
	\endfirsthead
		
	\multicolumn{6}{c}%
	{{\bfseries \tablename\ \thetable{} -- continued from previous page}} \\
	\hline \multicolumn{1}{|c|}{$M_{1} [\Ms]$} & \multicolumn{1}{c|}{$M_{2} [\Ms]$} & \multicolumn{1}{c|}{a [$R_{\sun}$]} & \multicolumn{1}{c|}{$Q_{t}$ } & \multicolumn{1}{c|}{$Q_{d}$ } & \multicolumn{1}{c|}{$Q_{c}$ }\\ \hline 
	\endhead
		
	\hline \multicolumn{6}{|r|}{{Continued on next page}} \\ \hline
	\endfoot
		
	\hline \hline
	\endlastfoot
		
	10&8&100&$8.59\times10^{60}$&$6.38\times10^{60}$&$2.21\times10^{60}$\\
	\hline
	10&5&900&$6.45\times10^{60}$&$6.38\times10^{60}$&$6.90\times10^{58}$\\
        10&5&1000&$6.45\times10^{60}$&$6.38\times10^{60}$&$6.90\times10^{58}$\\
	\hline
	20&16&100&$1.34\times10^{62}$&$8.15\times10^{61}$&$5.29\times10^{61}$\\
        20&16&200&$1.11\times10^{62}$&$7.80\times10^{61}$&$3.26\times10^{61}$\\
        20&16&300&$1.10\times10^{62}$&$7.77\times10^{61}$&$3.27\times10^{61}$\\
        20&16&400&$1.10\times10^{62}$&$7.77\times10^{61}$&$3.27\times10^{61}$\\
        20&16&500&$1.10\times10^{62}$&$7.77\times10^{61}$&$3.27\times10^{61}$\\
\end{longtable}
\end{center}

\begin{center}
\begin{longtable}{ |c|c|c|c|c|c| }
	\hline
	\multicolumn{6}{|c|}{Z = 0.01 \Zs, He II} \\
	\hline $M_{1} [\Ms]$& $M_{2} [\Ms]$ &a [$R_{\sun}$]&$Q_{t}$ &$Q_{d}$ &$Q_{c}$ \\ \hline
	\endfirsthead
		
	\multicolumn{6}{c}%
	{{\bfseries \tablename\ \thetable{} -- continued from previous page}} \\
	\hline \multicolumn{1}{|c|}{$M_{1} [\Ms]$} & \multicolumn{1}{c|}{$M_{2} [\Ms]$} & \multicolumn{1}{c|}{a [$R_{\sun}$]} & \multicolumn{1}{c|}{$Q_{t}$ } & \multicolumn{1}{c|}{$Q_{d}$ } & \multicolumn{1}{c|}{$Q_{c}$ }\\ \hline 
	\endhead
		
	\hline \multicolumn{6}{|r|}{{Continued on next page}} \\ \hline
	\endfoot
		
	\hline \hline
	\endlastfoot
		
	10&8&100&$1.08\times10^{57}$&$3.06\times10^{56}$&$7.74\times10^{56}$\\
	\hline
	10&5&900&$3.06\times10^{56}$&$3.06\times10^{56}$&$7.89\times10^{52}$\\
        10&5&1000&$3.06\times10^{56}$&$3.06\times10^{56}$&$7.80\times10^{52}$\\
	\hline
	20&16&100&$1.06\times10^{59}$&$5.08\times10^{58}$&$5.47\times10^{58}$\\
        20&16&200&$6.12\times10^{58}$&$4.98\times10^{58}$&$1.14\times10^{58}$\\
        20&16&300&$6.11\times10^{58}$&$4.97\times10^{58}$&$1.14\times10^{58}$\\
        20&16&400&$6.11\times10^{58}$&$4.97\times10^{58}$&$1.14\times10^{58}$\\
        20&16&500&$6.11\times10^{58}$&$4.97\times10^{58}$&$1.14\times10^{58}$\\
\end{longtable}
\end{center}

\begin{center}
\begin{longtable}{ |c|c|c|c|c|c| }
	\hline
	\multicolumn{6}{|c|}{Z = 0.01 \Zs, LW} \\
	\hline $M_{1} [\Ms]$& $M_{2} [\Ms]$ &a [$R_{\sun}$]&$Q_{t}$ &$Q_{d}$ &$Q_{c}$ \\ \hline
	\endfirsthead
		
	\multicolumn{6}{c}%
	{{\bfseries \tablename\ \thetable{} -- continued from previous page}} \\
	\hline \multicolumn{1}{|c|}{$M_{1} [\Ms]$} & \multicolumn{1}{c|}{$M_{2} [\Ms]$} & \multicolumn{1}{c|}{a [$R_{\sun}$]} & \multicolumn{1}{c|}{$Q_{t}$ } & \multicolumn{1}{c|}{$Q_{d}$ } & \multicolumn{1}{c|}{$Q_{c}$ }\\ \hline 
	\endhead
		
	\hline \multicolumn{6}{|r|}{{Continued on next page}} \\ \hline
	\endfoot
		
	\hline \hline
	\endlastfoot
		
	10&8&100&$2.29\times10^{62}$&$1.59\times10^{62}$&$7.00\times10^{61}$\\
	\hline
	10&5&900&$1.69\times10^{62}$&$1.59\times10^{62}$&$9.55\times10^{60}$\\
        10&5&1000&$1.69\times10^{62}$&$1.59\times10^{62}$&$9.55\times10^{60}$\\
	\hline
	20&16&100&$7.84\times10^{62}$&$4.97\times10^{62}$&$2.87\times10^{62}$\\
        20&16&200&$7.33\times10^{62}$&$4.89\times10^{62}$&$2.44\times10^{62}$\\
        20&16&300&$7.29\times10^{62}$&$4.85\times10^{62}$&$2.44\times10^{62}$\\
        20&16&400&$7.29\times10^{62}$&$4.85\times10^{62}$&$2.44\times10^{62}$\\
        20&16&500&$7.29\times10^{62}$&$4.85\times10^{62}$&$2.44\times10^{62}$\\
\end{longtable}
\end{center}

\begin{center}
\begin{longtable}{ |c|c|c|c|c|c| }
	\hline
	\multicolumn{6}{|c|}{Single star} \\
	\hline $M [\Ms]$& $Z [\Zs]$ & H I& He I& He II& LW \\ \hline
	\endfirsthead
		
	\multicolumn{6}{c}%
	{{\bfseries \tablename\ \thetable{} -- continued from previous page}} \\
	\hline \multicolumn{1}{|c|}{$M [\Ms]$} & \multicolumn{1}{c|}{$Z [\Zs]$} & \multicolumn{1}{c|}{H I} & \multicolumn{1}{c|}{He I} & \multicolumn{1}{c|}{He II } & \multicolumn{1}{c|}{LW }\\ \hline 
	\endhead
		
	\hline \multicolumn{6}{|r|}{{Continued on next page}} \\ \hline
	\endfoot

        \hline \hline
	\endlastfoot

        10&0&$3.66\times10^{62}$&$5.37\times10^{61}$&$1.02\times10^{59}$&$1.59\times10^{62}$\\
        20&0&$1.15\times10^{63}$&$2.94\times10^{62}$&$3.26\times10^{60}$&$3.28\times10^{62}$\\
        30&0&$2.29\times10^{63}$&$6.98\times10^{62}$&$1.37\times10^{61}$&$5.62\times10^{62}$\\
        40&0&$3.53\times10^{63}$&$1.18\times10^{63}$&$3.16\times10^{61}$&$8.03\times10^{62}$\\
        50&0&$4.82\times10^{63}$&$1.69\times10^{63}$&$5.53\times10^{61}$&$1.05\times10^{63}$\\
        60&0&$6.13\times10^{63}$&$2.22\times10^{63}$&$8.32\times10^{61}$&$1.30\times10^{63}$\\
        70&0&$7.43\times10^{63}$&$2.74\times10^{63}$&$1.14\times10^{62}$&$1.58\times10^{63}$\\
        80&0&$9.48\times10^{63}$&$3.46\times10^{63}$&$1.49\times10^{62}$&$2.04\times10^{63}$\\
\end{longtable}
\end{center}

\end{document}